\documentclass[journal]{IEEEtran}

\usepackage[T1]{fontenc}
\usepackage[utf8]{inputenc}
\usepackage{amsmath,amssymb,bm}
\usepackage{graphicx}
\usepackage{booktabs}
\usepackage{array}
\usepackage{siunitx}
\usepackage[numbers,sort&compress]{natbib}
\usepackage{url}
\usepackage{xcolor}
\usepackage[colorlinks=true,allcolors=blue!55!black]{hyperref}

\newcommand{\snot}{\sigma^{0}}
\newcommand{\mss}{\mathrm{mss}}
\newcommand{\msseff}{\mathrm{mss}_{\mathrm{eff}}}
\newcommand{\Uten}{U_{10}}
\newcommand{\Hs}{H_s}
\newcommand{\dB}{\,\mathrm{dB}}
\newcommand{\qz}{q_z}
\newcommand{\qh}{q_H}
\newcommand{\rstar}{r^{*}}
\newcommand{\Dtwo}{D_2}
\newcommand{\dmss}{\Delta\mathrm{mss_{shape}}}

\graphicspath{{figures/}}

\begin{document}
\bstctlcite{IEEEexample:BSTcontrol}

\title{A Dual-Frequency-Constrained Structure-Function\\
Kirchhoff Model for Near-Nadir Ocean Backscatter\\
---Part~II: The Practical GMF, KuKa-Mod}

\author{Alexey~S.~Mironov,
        Fr\'ed\'eric~Nouguier,
        and~Bertrand~Chapron%
\thanks{A.~S.~Mironov is with eOdyn, Plouzan\'e, France
        (e-mail: alexey.mironov@eodyn.com).}%
\thanks{F.~Nouguier and B.~Chapron are with the Laboratoire d'Oc\'eanographie
        Physique et Spatiale (LOPS), IFREMER, Plouzan\'e, France
        (e-mail: frederic.nouguier@ifremer.fr; bchapron@ifremer.fr).}%
\thanks{This work has been submitted to the IEEE for possible publication. Copyright may be transferred without notice, after which this version may no longer be accessible.}}

\markboth{Preprint submitted to IEEE Transactions on Geoscience and Remote Sensing}%
{Mironov \emph{et al.}: Structure-Function Kirchhoff Model for Near-Nadir Backscatter --- Part II}

\maketitle

\begin{abstract}
Near-nadir radars, from altimeters and rain radars to wave scatterometers, need a geophysical model
function (GMF) relating backscatter to wind, yet most operational models are empirical and specific to one band.
This paper delivers KuKa-Mod, an operational near-nadir GMF for wind retrieval at Ku and Ka band, built
on one physical description of the sea surface, and evaluates it. Following Part~I, the additive coefficient and log-slope of a logarithmic surface
structure function are fitted jointly to the Ku-band incidence profile of the Global Precipitation
Measurement (GPM) radar and to the same-footprint Ku--Ka shape difference. The physical core reproduces the Ku
profile and reaches the azimuth-averaged Ka profile with no additional shape parameter and one fitted level, largely explained by
the seawater dielectric and the clear-sky attenuation. Against scatterometer winds the Ka--Ku difference
holds to within $0.3\dB$ up to $25\,\mathrm{m\,s^{-1}}$, whereas the near-nadir Ka wind sensitivity is about
$20\%$ too weak. An empirical correction sets the operational level of both bands, with held-out quantile
agreement over $8$--$20\,\mathrm{m\,s^{-1}}$ in the tested near-nadir bins. Without a sea-state coefficient, a long-wave
tilt predicts where the response to significant wave height changes sign at Ku band, a tilt that
CFOSAT/SWIM qualitatively supports, and the azimuth harmonics are reproduced once a sensor-dependent offset is removed.
The physical core is released as parameters and an evaluator, and the operational levels as
look-up tables.
\end{abstract}


\section*{Nomenclature}
\begin{IEEEdescription}[\IEEEsetlabelwidth{$\Delta\mathrm{mss_{shape}}$}]
\item[$\snot$] normalized radar cross-section (linear unless in dB).
\item[$\Dtwo(r)$] second structure function, the mean-square elevation difference $\langle[\eta(\bm x+\bm r)-\eta(\bm x)]^2\rangle$ at lag $r$.
\item[$s^2(r)$] lag-dependent slope variance $2\Dtwo(r)/r^2$.
\item[$a,b$] additive coefficient and log-slope of $s^2(r)$ in \eqref{eq:marginallog}, fitted at six wind nodes.
\item[$r_{\rm ref}$] reference lag, fixed at $1$\,m.
\item[$\qz,\qh$] vertical and horizontal components of the scattering vector.
\item[$\rstar$] kernel e-folding lag, $\Dtwo(\rstar)=2/\qz^2$.
\item[mss] mean-square slope.
\item[$\msseff$] radar-filtered mean-square slope, $s^2(\rstar)$.
\item[$\dmss$] Ka minus Ku shape mss from the regression \eqref{eq:shapemss}.
\item[msc] mean-square curvature ($\mathrm{m^{-2}}$), the curvature term of GO4.
\item[$R_{\rm eff}^2$] nadir level factor $R_F^2(f,T,S)\,C_f$.
\item[$\Phi$] wind-relative azimuth, $\Phi=0$ looking downwind.
\item[$A_n,a_n$] azimuth harmonics of $\snot$, linear and in dB, $a_n\simeq(10/\ln10)A_n$.
\item[$D_{20}$] isotropic part of $\Dtwo(r,\psi)$.
\item[$D_{22}$] $\cos2\psi$ part of $\Dtwo(r,\psi)$, with coefficients $\delta_a,\delta_b$.
\item[$\lambda_3^{0},D_c$] amplitude and crossover of the odd closure \eqref{eq:full_dir}.
\item[$\Delta\sigma^2_{\rm long}$] long-wave slope variance in excess of its value at the reference sea state \eqref{eq:hsmss}.
\item[hinge] incidence where $\tan^2\theta=\msseff$, at which the GO cross-section is insensitive to wind and to $\Hs$.
\item[$\sigma_U$] within-footprint wind standard deviation.
\item[$U_{10S}$] stress-equivalent $10$-m wind of the scatterometers.
\item[$\Delta_{\rm hw}$] empirical level correction of the high-wind tables (dB).
\item[PIA] two-way path-integrated attenuation.
\end{IEEEdescription}

\section{Introduction}
\label{sec:intro}

Wind speed over the ocean is routinely retrieved from radar backscatter measured
near vertical incidence. Altimeters, the precipitation radars of the Tropical
Rainfall Measuring Mission (TRMM) and the Global Precipitation Measurement (GPM)
mission, and the Surface Waves Investigation and Monitoring (SWIM) wave scatterometer on the
China--France Oceanography
Satellite (CFOSAT) all record this return
\citep{chelton1986,meneghini2000,freilich2003,hauser2021}. Each retrieval inverts
a geophysical model function (GMF), which gives the normalized radar cross-section
$\snot$ as a function of wind, incidence, azimuth and sea state. A wind processor
needs this GMF in tabulated form, referenced to a stable absolute calibration and
with a stated validity envelope.

Three practical needs are not met by the GMFs in use. First, near-nadir
measurements are increasingly made in two bands. The GPM radar observes each
footprint at Ku and Ka band, Ka band is carried by the AltiKa altimeter and the
Surface Water and Ocean Topography (SWOT) swath altimeter, and the planned
Copernicus polar ice and snow topography altimeter (CRISTAL) combines both
\citep{hou2014,verron2015,morrow2019,kern2020}. Inter-calibrating these missions,
and retrieving wind from a Ka-only instrument, require one GMF that serves both
bands consistently. Current near-nadir GMFs, however, are empirical fits made one
band and one instrument at a time \citep{gourrion2002,lillibridge2014,ren2016,hossan2021}, so a
new Ka mission cannot build on the long Ku record. Second, near-nadir platforms
often measure the significant wave height $\Hs$ together with $\snot$, yet
altimeter algorithms include it only empirically \citep{gourrion2002}, the
published precipitation-radar and off-nadir GMFs omit it \citep{hossan2021,ren2021},
and its effect on a near-nadir retrieval has not been quantified.
Third, a rotating-beam instrument such as SWIM samples the full wind-relative
azimuth and needs an azimuth term that is consistent with the incidence profile.

Part~I of this paper \citep{part1} provides the physical basis for such a GMF. It
writes the near-nadir Kirchhoff integral through the structure function of the sea
surface, the mean-square elevation difference between two surface points a
horizontal lag apart. The radar then selects a centimetre-scale lag, different
for each band, so Ku and Ka band read the same structure function at two lags.
With a logarithmic structure function, a level and a log-slope set both the
fall-off of $\snot$ with incidence and the Ku--Ka difference. Part~I also shows
that one band constrains these two parameters very unequally, so they must be
retrieved from matched dual-frequency observations.

The present paper performs that retrieval and turns it into an operational GMF,
KuKa-Mod. GPM observes the same ocean footprint simultaneously at Ku and Ka band,
and $1.6\times10^9$ such matched pairs, together with the Ku incidence profile,
constrain the coefficient and log-slope from $4$ to $16\,\mathrm{m\,s^{-1}}$. The resulting model is evaluated against
GPM and CFOSAT/SWIM, extended with an azimuth term, tested against the measured
sea-state response, and
released as a parameter set, a reference evaluator and look-up tables. It is compared
with the fourth-order geometric-optics model (GO4) of \citet{nouguier2009phd} and
\citet{boisot2015}, fitted on the same footing.

Sections~\ref{sec:model}--\ref{sec:fit} outline the model, the data and the fit.
Sections~\ref{sec:val} and \ref{sec:axes} evaluate it and give its high-wind level,
Section~\ref{sec:discussion} compares it with prior models and gives its operational
use, and Section~\ref{sec:disc-limits} states its limits.

\section{The KuKa-Mod Model: Outline of Part~I}
\label{sec:model}

KuKa-Mod combines a specular core, azimuth harmonics and a Bragg flank.
Part~I derives the core and the directional closures.

\subsection{Specular core and kernel lag}
Below about $12^\circ$ incidence the return is dominated by specular reflection from
surface facets oriented perpendicular to the beam. Part~I writes the Kirchhoff
integral through the second structure function
$\Dtwo(r)=\langle[\eta(\bm x+\bm r)-\eta(\bm x)]^2\rangle$ of the surface elevation
$\eta$. For a Gaussian, isotropic elevation increment the azimuth-averaged
cross-section reads
\begin{equation}
  \bar\snot(\theta)=\frac{2k_r^2 R_{\rm eff}^2}{\cos^2\theta}
    \int_0^\infty r\,J_0(\qh r)\,e^{-\qz^2\Dtwo(r)/2}\,dr,
  \label{eq:hankel}
\end{equation}
where $J_0$ is the Bessel function of order zero, $k_r$ is the radar wavenumber,
$\qz=2k_r\cos\theta$ and $\qh=2k_r\sin\theta$
are the vertical and horizontal components of the scattering vector, and
$R_{\rm eff}^2$ is the nadir level factor. When $\Dtwo$ is quadratic,
$\Dtwo=\tfrac12\mss\,r^2$, \eqref{eq:hankel} reduces to Gaussian, isotropic geometric optics (GO),
\begin{equation}
  \snot_{\rm GO}(\theta)=\frac{R_{\rm eff}^2}{\mss\cos^4\theta}
    \exp\!\Big(-\frac{\tan^2\theta}{\mss}\Big),
  \label{eq:go}
\end{equation}
with mss the mean-square slope. The kernel of \eqref{eq:hankel} weights a finite range
of lags around the e-folding lag $\rstar$, defined by
\begin{equation}
  \Dtwo(\rstar)=\frac{2}{\qz^2},
  \qquad
  \msseff\equiv s^2(\rstar)=\frac{2\Dtwo(\rstar)}{{\rstar}^2}.
  \label{eq:rstar}
\end{equation}
The slope variance at that lag, $\msseff$, is the radar-filtered mean-square slope
\citep{jackson1992,vandemark2004}, here obtained without an imposed spectral
cutoff. Over $3$--$20\,\mathrm{m\,s^{-1}}$, $\rstar$ falls from $2.5$ to $1.4$\,cm
at Ku band and from $1.0$ to $0.5$\,cm at Ka band. The two bands therefore read one
$\Dtwo(r)$ at two centimetre-scale lags.

\subsection{Logarithmic structure function and band transfer}
At these lags the saturation spectrum is nearly constant \citep{phillips1958}
and adds equal slope variance per logarithmic wavenumber interval, so
$s^2(r)=2\Dtwo(r)/r^2$ varies with lag only logarithmically. Part~I adopts
\begin{equation}
  \Dtwo(r)=\tfrac12 r^2\,s^2(r),\qquad
  s^2(r)=a+b\,\ln\!\Big(1+\frac{r_{\rm ref}}{r}\Big),
  \label{eq:marginallog}
\end{equation}
with the reference lag fixed by convention at $r_{\rm ref}=1$\,m. The log-slope
$b$ is the saturation level of the short waves, and $a$ is an additive level
coefficient, with $s^2(r_{\rm ref})=a+b\ln2$. Read at the two kernel lags, this form gives the band-transfer law
\begin{equation}
  \msseff(\lambda)\simeq\mathrm{const}-b\,\ln(\lambda/\lambda_0)
  \label{eq:bandlaw}
\end{equation}
to leading order in $b/\msseff$, where $\lambda$ is the radar wavelength. The same
$b$ thus sets the fall-off of $\snot$ with incidence and the Ku--Ka difference.
Each band, however, supplies essentially one effective slope variance, whereas
$s^2(r)$ has two unknowns. One band therefore fixes mainly the level and leaves $b$
poorly determined, and the log-slope must be retrieved from two bands. This is the
sense in which the model is dual-frequency-constrained.

\subsection{The complete model}
The physical core adds azimuth harmonics and an outer-incidence Bragg term to the
specular core (the operational tables also apply the level correction of
Section~\ref{sec:val-highwind}),
\begin{equation}
  \snot(\theta,\Uten,\Phi)
  =\bar\snot(\theta,\Uten)\big[1 + A_1\cos\Phi + A_2\cos2\Phi\big]
   + \snot_{\rm Bragg},
  \label{eq:full}
\end{equation}
where $\Phi$ is the wind-relative azimuth ($\Phi=0$ looking downwind) and
$\bar\snot$ is \eqref{eq:hankel} with $a(\Uten)$ and $b(\Uten)$ fitted at six wind
nodes (Section~\ref{sec:fit}). The harmonics describe the up/crosswind anisotropy
of $\Dtwo$ and the upwind/downwind asymmetry carried by the third
elevation-increment cumulant $D_3$ (Part~I),
\begin{equation}
  \begin{aligned}
  \Dtwo(r,\psi)&=D_{20}(r)+D_{22}(r)\cos2\psi,\\
  D_3(r,\psi)&=\lambda_3^{0}\,\frac{D_{20}^{3/2}}{\sqrt{1+D_{20}/D_c}}\,\cos\psi,
  \end{aligned}
  \label{eq:full_dir}
\end{equation}
where $\psi$ is the lag azimuth relative to the wind, $D_{20}$ is the isotropic
form \eqref{eq:marginallog}, $D_{22}=\tfrac12 r^2[\delta_a+\delta_b\ln(1+r_{\rm
ref}/r)]$, and $D_c=10^{-7}\,\mathrm{m^2}$ sets the crossover from $D_{20}^{3/2}$ to $D_{20}$
scaling. The odd term $D_3$ gives the upwind/downwind harmonic $A_1$, and
$D_{22}$ the up/crosswind harmonic $A_2$. In
decibels, $a_n\simeq(10/\ln10)A_n$. The fitted odd closure can represent variance--slope
correlations across scales and does not identify a unique modulation mechanism.

Beyond about $11^\circ$ an empirical Bragg term supplements the specular return and
extends the model to $18^\circ$,
\begin{equation}
  \snot_{\rm Bragg}=
  g(\theta)\,\beta\,(\Uten/10)^{\xi}\,
  e^{\,\alpha(\theta-14^\circ)/4^\circ}\,\snot_{\rm SPM}(\theta,\Uten),
  \label{eq:full_bragg}
\end{equation}
where $\snot_{\rm SPM}$ is the first-order small-perturbation (Bragg)
cross-section for horizontal transmit and receive (HH) polarization, the gate $g=\min\{1,\max[0,12\sin\theta-2]\}$
rises from zero at $9.6^\circ$ to one at $14.5^\circ$, and $(\beta,\xi,\alpha)$ are
fitted on Ku band. The term is negligible below $11^\circ$, and the Ka flank
inherits the Ku ramp.

The band enters through $k_r$ and through the level factor
$R_{\rm eff}^2=R_F^2(f,T,S)\,C_f$. Here $R_F^2$ is the nadir Fresnel reflectivity
computed from the seawater permittivity of \citet{meissner2004} at frequency $f$,
temperature $T$ and salinity $S$, and $C_f$ is a per-band calibration and path
factor that also absorbs the loss from unresolved facet roughness (Part~I). Sea state enters as an additive long-wave tilt $\Delta\sigma^2_{\rm long}(\Uten,\Hs)$ of the coefficient $a$ (Part~I). The
released model is evaluated at the reference sea state, and
Section~\ref{sec:val-hs} tests the tilt against SWIM and GPM. Part~I also gives a non-Gaussian (kurtosis) closure of
the increment, but two frequencies do not fix it, so the KuKa-Mod core has no
independently fitted kurtosis, although variance mixing still produces one (Part~I). The GMF is fitted and delivered at the GPM
footprint scale of about $5$\,km, and other footprints are reached by the
sub-footprint correction of Section~\ref{sec:val-inc}.

\section{Data and Processing}
\label{sec:data}

KuKa-Mod is fitted and evaluated on near-nadir backscatter from two missions. GPM
fixes the absolute level and both bands, while CFOSAT/SWIM adds dense azimuth
sampling and a closely collocated sea state.

\subsection{GPM dual-frequency precipitation radar}
\label{sec:data-gpm}
The primary data are the ocean cross-sections of the Dual-frequency Precipitation
Radar (DPR) on the GPM core observatory \citep{hou2014,seto2021}, which observes
each scene simultaneously at Ku band ($13.6$\,GHz, KuPR) and Ka band
($35.5$\,GHz, KaPR). We use the V07A product over $0$--$18^\circ$ incidence,
binned in $0.25^\circ$ incidence and $0.5\,\mathrm{m\,s^{-1}}$ wind over
$3$--$20\,\mathrm{m\,s^{-1}}$.
Two reductions are used. The KuPR incidence profile, to which the model is fitted,
is the 2021 reduction at the reference sea state $\Hs\in[1.5,3.5]$\,m ($12$ months, $902$ populated bins, wind-bin centres up to
$14.25\,\mathrm{m\,s^{-1}}$). The shape difference is built from 2021-12 to 2023-07
($19$ months with usable data), without $\Hs$ stratification. KuPR is the single inter-calibration reference:
every other sensor and published GMF is re-referenced to it by a measured constant
offset.

The shape difference is measured on pairs joined by orbit, scan and ray index, so
that the two frequencies share the same footprint to within $0.64$\,km. Such
same-footprint pairs are called \emph{matched} pairs. In their difference the
common long-wave contribution is strongly suppressed. The $1.6\times10^9$ matched
pairs are reduced per incidence/wind bin to the per-band shape mss, the slope variance
inferred from the incidence fall-off (Section~\ref{sec:proc-shape}), and to its shape difference $\dmss$, a direct probe
of $\Dtwo$ at centimetre lags.

\subsection{CFOSAT/SWIM}
\label{sec:data-swim}
SWIM is a Ku-band ($13.575$\,GHz) real-aperture wave scatterometer whose antenna
rotates in azimuth while illuminating beams at nominal $0$, $2$, $4$, $6$, $8$ and
$10^\circ$ \citep{hauser2017,hauser2021}. Each off-nadir beam thus samples the full
$360^\circ$ of wind-relative azimuth. We use the L2S product (v2.0) for the five
off-nadir beams. The nadir beam measures $\Hs$ on the same ground track, which
gives a far closer sea-state collocation than a model field. SWIM serves as an
auxiliary, not an absolute, reference. As a wave scatterometer its absolute level
is not maintained to scatterometric standards, and each beam has its own antenna
gain. SWIM $\snot$ is therefore recalibrated per beam to KuPR (Appendix~\ref{app:proc}).
The recalibration acts on the level and in the orbital frame, so it is designed to
leave the wind-relative harmonics unchanged, which the SWIM--GPM agreement of the
first harmonic supports (Section~\ref{sec:val-dir}).

\subsection{Auxiliary models and reference GMFs}
\label{sec:data-models}
The ERA5 reanalysis \citep{hersbach2020} provides the collocated $10$-m wind
$\Uten$, to which the released GMF is referenced, and the significant wave height.
Scatterometer winds give a second, stress-equivalent reference $U_{10S}$
\citep{dekloe2017}. They are the level-2 winds of the Advanced Scatterometer (ASCAT) on
Metop-A/B/C, HY-2B/C and ScatSat-1 from the Royal Netherlands Meteorological
Institute (KNMI), taken from the pass closest in time within $25$\,km and $90$\,min
($60$\,min for GPM, $30$\,min for the SWIM triple collocation), and shifted to the
radar time by the change of the ERA5 wind between the two times. For GPM, all KuPR and
KaPR footprints from December 2021 to July 2023 with an ERA5 wind of at least
$12\,\mathrm{m\,s^{-1}}$ and a $1\%$ random subsample of the others are kept, with
inverse-probability weights, and Ka and Ku footprints within $1.5$\,km and $5$\,s are
paired. The scatterometer winds serve Sections~\ref{sec:val-stab}--\ref{sec:val-highwind}
and \ref{sec:val-hs}. ERA5 is a
reference rather than the truth. Its speed errors attenuate the fitted wind
dependence, and its direction errors attenuate the azimuth harmonics, the second
more than the first \citep{belmonte2019}. WAVEWATCH~III \citep{ww3dg2019} supplies
the wind-sea/swell partition. The short-wave spectrum of \citet{kudryavtsev2014}
(hereafter KMC), with its wave-breaking and parasitic-capillary terms, is the physical spectrum against which the fitted log-slope is
compared.

The model is benchmarked against established near-nadir models, each
re-referenced to KuPR so that only the shape is compared. The equal-parameter
baseline is geometric optics with the same Bragg flank and a band-invariant
$c_{40}$, the fourth-order peakedness coefficient of a Gram--Charlier slope probability density
(hereafter the Gram--Charlier control). The GO4
model \citep{nouguier2009phd,boisot2015}, which adds a curvature term, is fitted
separately (Section~\ref{sec:disc-go}). The empirical band-difference relation of
\citet{nouguier2016} serves as the reference for $\dmss$. The \citet{hossan2021}
DPR GMF is re-referenced by its measured $+0.32\dB$ offset, and the KuLMOD SWIM
power law \citep{ren2021} is refitted.

\subsection{Shape and harmonic estimators}
\label{sec:proc-shape}
In each band the azimuth-averaged $\snot$ is averaged in dB per $(\theta,\Uten)$
bin and fitted with $\sqrt{n}$ count weighting, which approximates inverse-variance
weighting. The core fit targets the dB mean, so the look-up tables give the
dB-mean cross-section converted to linear units; it differs from the linear mean
by a median $0.044\dB$ over the $902$ bins. Following \citet{nouguier2016}, the
$2$--$9^\circ$ fall-off of $\snot$ with incidence defines a per-band shape mss,
\begin{equation}
  10\log_{10}(\cos^4\theta\,\snot)
   = \mathrm{const}-\frac{10}{\ln10}\frac{\tan^2\theta}{\mathrm{mss_{shape}}}.
  \label{eq:shapemss}
\end{equation}
The shape difference $\dmss=\mathrm{mss_{shape}^{Ka}}-\mathrm{mss_{shape}^{Ku}}$
removes most of the common long-wave contribution and isolates $\Dtwo$ at
centimetre lags. Model and data pass through the same regression. The SWIM
harmonics are fitted per beam and wind class as
$\snot_{\rm dB}=a_0+a_1\cos\Phi+a_2\cos2\Phi$ relative to the ERA5 wind direction.

\subsection{Uncertainty}
\label{sec:proc-boot}
Observations within one orbit share one geophysical state, so the binned
statistics hold far fewer independent degrees of freedom than their counts
suggest. Sampling uncertainties are therefore taken from a calendar-month block
bootstrap of the estimator actually used. The $19$ monthly matched reductions, and
in a second variant the $12$ KuPR months, are resampled with replacement, and the
full joint model is refitted on each of $300$ replicates. The
block standard error of the per-band shape mss is some three orders of magnitude
above the naive per-bin error. Yet the matched difference cancels the common
long-wave variability, so $\dmss$ is stable across months to $\pm1$--$4\%$ per
wind bin, and so are the fitted $b$ nodes. Fitted as $b\propto\Uten^{\,p}$ over
$5$--$16\,\mathrm{m\,s^{-1}}$, the wind exponent is $p=2.00$, with a $95\%$
month-block bootstrap interval of $1.98$--$2.02$. The fit settings matter more than sampling: the objective
weights move $p$ between $1.45$ and $2.06$.

Sampling is not the largest error. Each error source was injected into the joint
inputs and the model refitted. A constant Ku calibration offset, a constant path
attenuation or the dielectric level is absorbed by $R_{\rm eff}^2$, and constant Ka
offsets cancel from the shape regression. The most sensitive term is an incidence
trend of the Ka path-integrated attenuation (PIA), because $\dmss$ is formed from
uncorrected cross-sections. A worst-case trend of $0.2\dB$ per $10^\circ$ would move
$b$ by up to $94\%$, but the measured trend is below $0.01\dB$
(Section~\ref{sec:val-band}), which reduces the effect to about $5\%$. Wind errors of
$1\,\mathrm{m\,s^{-1}}$ r.m.s.\ move $b$ by up to $23\%$, mostly at
$20\,\mathrm{m\,s^{-1}}$ where the fit is extrapolated. An incidence
misregistration of $\pm0.1^\circ$ moves $b$ by less than $2\%$. Halving the weight $w_\Delta$
of the shape difference lets $b$ drift back toward the inflated one-band solution, so
it is set large enough to keep that constraint active.

\section{Constructing the GMF: The Joint Retrieval}
\label{sec:fit}

The KuPR profile and the matched shape difference jointly constrain the level
$a(\Uten)$ and the log-slope $b(\Uten)$ of the GMF. Table~\ref{tab:exp} separates that fit from the
diagnostic fits and the withheld-band test. Row~4, with the Ka data
withheld, is the table's only out-of-sample band test. Because the physical-core $b(\Uten)$ is fitted to the shape
difference, the cross-band agreement of Section~\ref{sec:val} is otherwise an
internal-consistency check.

\begin{table*}[t]
  \centering
  \caption{Provenance of the reported results (tests in-sample unless marked
  \textbf{withheld}). The ERA5 physical core, row~3, has $16$ parameters: $6$ $a$
  nodes, $6$ $b$ nodes, the level factor $R_{\rm eff}^2$ and three Bragg-flank
  parameters. wRMSD: $\sqrt n$-weighted root-mean-square difference.}
  \label{tab:exp}
  \footnotesize
  \setlength{\tabcolsep}{4pt}
  \renewcommand{\arraystretch}{1.25}
  \begin{tabular}{@{}p{0.05\textwidth}p{0.26\textwidth}p{0.25\textwidth}p{0.22\textwidth}p{0.10\textwidth}@{}}
    \toprule
    \# & Model variant & Fitted to & Test / reported & Free params \\
    \midrule
    1 & logarithmic $\Dtwo$, wind in $b(\Uten)$ only
      & KuPR profile, $0$--$18^\circ$
      & in-sample wRMSD $0.116\dB$
      & 11 \\
    2 & KMC-spectrum drive (predictive)
      & KuPR profile
      & in-sample $0.113\dB$
      & 4 \\
    3 & logarithmic $a(\Uten),b(\Uten)$, Bragg flank (\textbf{physical core})
      & KuPR profile and matched $\dmss$ (joint)
      & in-sample $0.105\dB$; $\dmss$ RMSD $3.1\times10^{-4}$
      & 16 \\
    4 & same form, $\dmss$ term removed (\textbf{withheld})
      & KuPR profile \textbf{only}
      & \textbf{withheld} $\dmss$; KaPR shape after one fitted level
      & 16 \\
    5 & row~3 read at the Ka wavenumber
      & $a,b(\Uten)$ from row~3; one Ka level
      & KaPR profile (shape not refitted)
      & 1 \\
    6 & directional closure ($D_3$, anisotropy)
      & form from SWIM; amplitudes from GPM-Ku
      & in-sample $a_1,a_2$ to $0.01$--$0.02\dB$
      & $\lambda_3^{0},\delta_a,\delta_b$ \\
    7 & GO4 (free $\mss$, msc), objective of row~3
      & KuPR profile and matched $\dmss$ (joint)
      & in-sample $0.100\dB$; $\dmss$ $4.9\times10^{-4}$; KaPR $0.26\dB$
      & 16 \\
    8 & $U_{10S}$ core refit (directional module unchanged)
      & KuPR--scatterometer 2021--2022, $1.69\times10^6$ samples
      & in-sample wRMSD $0.103\dB$; no separate holdout
      & 16 \\
    9 & operational tables: core $\times10^{\Delta_{\rm hw}/10}$
      & GPM--scatterometer 2021-12--2023-07, $1.02\times10^8$ Ku / $1.08\times10^8$ Ka
        footprints; odd months fitted
      & \textbf{withheld} even months: quantile residual $+0.02$ / $0.00\dB$ (Ku/Ka) above
        $12\,\mathrm{m\,s^{-1}}$; quantile-matched bins only ($\theta\lesssim6^\circ$ Ku,
        $7.5^\circ$ Ka), distributional assessment to $20\,\mathrm{m\,s^{-1}}$
      & empirical table \\
    \bottomrule
  \end{tabular}
\end{table*}

\subsection{The physical-core joint fit}
\label{sec:fit-joint}
Both $a$ and $b$ are free at six wind nodes ($3$, $5$, $8$, $12$, $16$ and
$20\,\mathrm{m\,s^{-1}}$), interpolated in $\ln\Uten$ by piecewise cubic Hermite
interpolation (PCHIP). They are fitted by minimizing
\begin{equation}
  \begin{aligned}
  J={}& \sum_{i}w_i^2\big[{\snot}_{\rm dB,mod}(\theta_i,U_{10,i})-{\snot}_{\rm dB,KuPR}\big]^2\\
    &{}+w_\Delta^2\!\sum_j\big[\dmss^{\rm mod}_j-\dmss^{\rm obs}_j\big]^2\\
    &{}+\lambda_s^2\big(\|\Delta^2 a\|^2+\|\Delta^2 b\|^2\big),
  \end{aligned}
  \label{eq:objective}
\end{equation}
where the profile weights $w_i=v_i/\langle v\rangle$, with
$v_i=\sqrt{\min(n_i,10^7)}$, follow the bin counts $n_i$. The weight $w_\Delta=500$ puts the $\mathcal O(10^{-3})$ band-difference
residual on the scale of the dB profile residual. The second-difference penalty
$\lambda_s=30$ controls the nodes above $16\,\mathrm{m\,s^{-1}}$, which neither the
KuPR profile (direct support to about $14\,\mathrm{m\,s^{-1}}$) nor the matched
$\dmss$ reaches. The level factor and the three Bragg parameters are fitted
together with the twelve nodes.

\begin{figure*}[t]
  \centering
  \includegraphics[width=\textwidth]{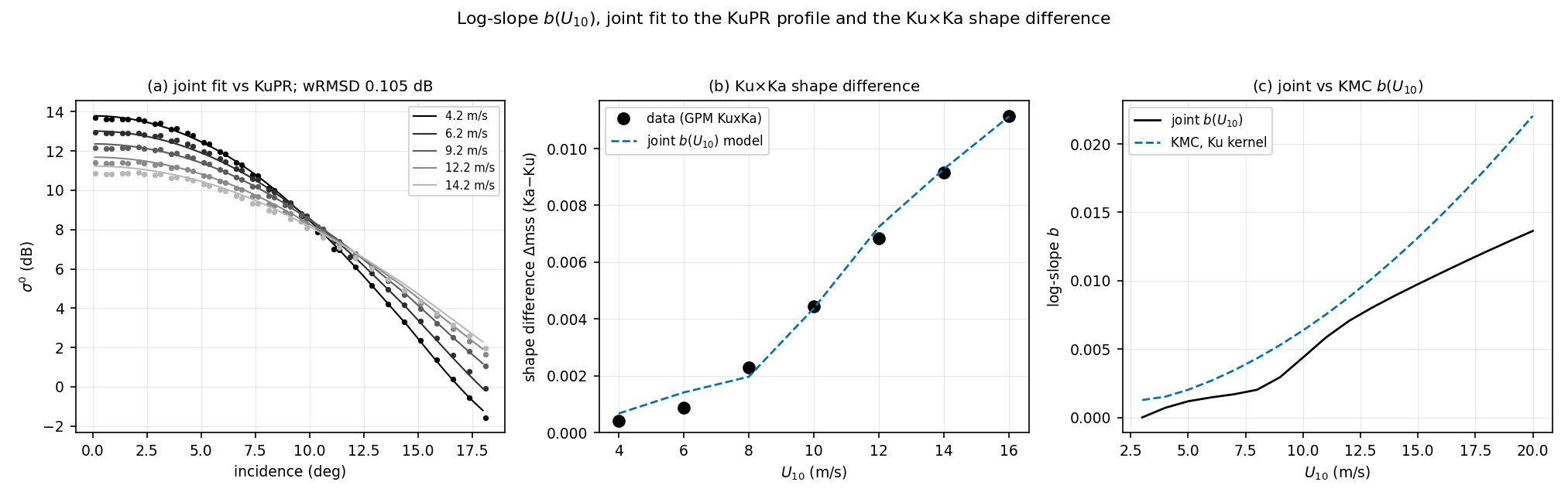}
  \caption{The physical-core joint fit. (a)~Model against the KuPR profile (markers;
  wRMSD $0.105\dB$ over $902$ bins). (b)~The shape difference $\dmss$ against the
  matched GPM Ku$\times$Ka data, matched from $8\,\mathrm{m\,s^{-1}}$ upward and
  over-predicted by about $60\%$ at $4$--$6\,\mathrm{m\,s^{-1}}$. (c)~The fitted
  log-slope $b(\Uten)$ and the KMC log-slope at the Ku kernel, which is $40\%$
  above the fitted joint $b$ (and $35\%$ above the shape-difference log-slope, Part~I) at $16\,\mathrm{m\,s^{-1}}$.}
  \label{fig:bujoint}
\end{figure*}

As shown in Fig.~\ref{fig:bujoint}, the fit reproduces the KuPR profile with a
wRMSD of $0.105\dB$ on $902$
bins, or $0.094\dB$ with the weights $w_i^2$ of the objective. It reproduces the shape difference with an RMSD of $3.1\times10^{-4}$, whereas
the KMC spectrum predicts it with an RMSD of $2.6\times10^{-3}$, too high at high
wind. The shape difference is matched to within $15\%$ at $8\,\mathrm{m\,s^{-1}}$
and within $7\%$ from $10$ to $16\,\mathrm{m\,s^{-1}}$. At the fitted target winds of $4$ and $6\,\mathrm{m\,s^{-1}}$, where
$\dmss<10^{-3}$, the fit over-predicts it by about $60\%$. The fitted nodes (Table~\ref{tab:abnodes}) show $b$ growing approximately
as $\Uten^{2}$, as expected for the saturation level of the short waves (Part~I).
The coefficient $a$ decreases above $8\,\mathrm{m\,s^{-1}}$. Since the logarithmic form is a local closure over
the centimetre lags both kernels sample, $a$ is not a metre-scale slope variance.

\begin{table}[t]
  \centering
  \caption{Fitted physical-core parameters, released with the code: additive coefficient $a$ and
  log-slope $b$ at the six wind nodes ($r_{\rm ref}=1$\,m). The level factor is
  $R_{\rm eff}^2=0.581$ and the Bragg triple is
  $(\beta,\xi,\alpha)=(0.545,-1.74,1.09)$. The nodes $a(20)$ and $b(3)$ sit on the
  lower bounds of the fit.}
  \label{tab:abnodes}
  \footnotesize
  \setlength{\tabcolsep}{4pt}
  \begin{tabular}{lcccccc}
    \toprule
    $\Uten$ ($\mathrm{m\,s^{-1}}$) & 3 & 5 & 8 & 12 & 16 & 20 \\
    \midrule
    $a\ (\times10^{-2})$ & 1.90 & 2.22 & 2.42 & 1.20 & 0.64 & 0.20 \\
    $b\ (\times10^{-3})$ & 0.00 & 1.18 & 2.02 & 7.06 & 10.6 & 13.6 \\
    \bottomrule
  \end{tabular}
\end{table}

\subsection{Why the log-slope carries the wind}
\label{sec:fit-bU}
Two diagnostic fits to the Ku profile alone explain this construction. In the
first (row~1), the wind is carried either by $b$ or by $a$, with the other held
constant and the same $11$ parameters as the Gram--Charlier control. With the wind
in $b$ the wRMSD is $0.116\dB$, against $0.144\dB$ with the wind in $a$ and
$0.137\dB$ for the control. The gain is largest on the $11$--$18^\circ$ flank
($0.112$ against $0.157\dB$), where the small-lag part of $\Dtwo$ controls the
integral. With six mss and six $c_{40}$ nodes, a $16$-parameter control reaches $0.110\dB$,
against $0.105\dB$ for the Ku-only logarithmic fit of Part~I. In the second (row~2), $\Dtwo$ is generated from the KMC spectrum, with
only the level and the Bragg flank fitted. This drive reaches $0.113\dB$, and its
fitted level is close to the computed Fresnel value. The generated $b$ has nearly
the right wind exponent ($p\simeq1.8$ at Ku) but too high a level. At $16\,\mathrm{m\,s^{-1}}$ it exceeds the log-slope of the shape
difference by $35\%$ (Ku) and $74\%$ (Ka) (Part~I). The log-slope is therefore
taken from the radar, and the spectrum serves only as a prior on its form.

\subsection{The dual-frequency constraint}
\label{sec:fit-dualfreq}
\begin{figure*}[t]
  \centering
  \includegraphics[width=\textwidth]{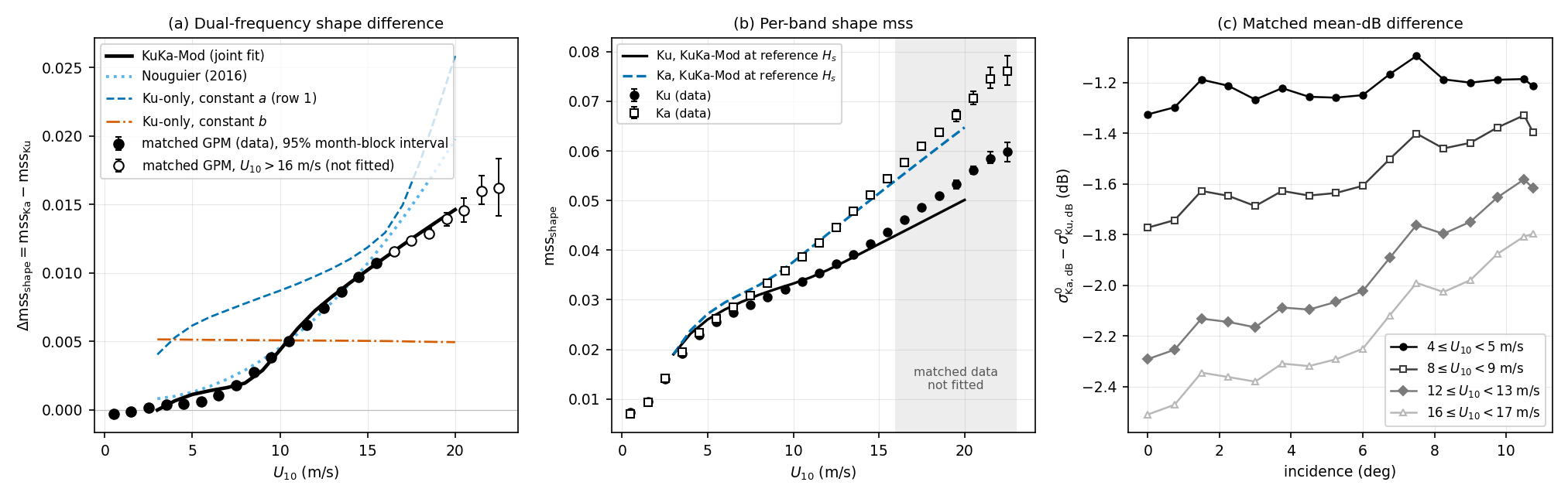}
  \caption{Dual-frequency shape difference on matched GPM Ku$\times$Ka.
  (a)~$\dmss$ against wind: KuKa-Mod (solid) matches the data (markers, month-block
  $95\%$ intervals) as closely as \citet{nouguier2016} (dotted). Ku-only fits with
  constant $a$ (dashed, Table~\ref{tab:exp}, row~1) over-predict at every wind,
  and with constant $b$ (dash-dot) stay near $0.005$. Data above
  $16\,\mathrm{m\,s^{-1}}$ are not fitted.
  (b)~Per-band shape mss, with KuKa-Mod at the reference sea state
  $\Hs\in[1.5,3.5]$\,m. It follows both bands within $10\%$ from $4$ to
  $8\,\mathrm{m\,s^{-1}}$ and $4\%$ from $8$ to $16\,\mathrm{m\,s^{-1}}$, and falls
  $4$--$8\%$ short above (shaded: not fitted), where the sampled $\Hs$ is well above
  the reference window. (c)~The matched mean-dB difference
  $\snot_{\rm Ka}-\snot_{\rm Ku}$, more negative at higher wind.}
  \label{fig:dualfreq}
\end{figure*}

The shape difference itself confirms that the log-slope carries the wind. As shown
in Fig.~\ref{fig:dualfreq}, $\dmss$ rises quadratically with wind. Over
$4$--$16\,\mathrm{m\,s^{-1}}$ its linear wind slope is
$+9.7\times10^{-4}$ per $\mathrm{m\,s^{-1}}$, against $+9.4\times10^{-4}$ for the
\citet{nouguier2016} relation on the same grid. Their relation is thus reproduced
on a later GPM period and product version. KuKa-Mod matches the measured rise
comparably, but through the same $b$ that sets the incidence fall-off, whereas the
empirical relation is a separate fit to the shape difference. With the wind in $a$
instead, the predicted Ka/Ku shape-mss ratio falls from $1.23$ to $1.10$ over
$4$--$16\,\mathrm{m\,s^{-1}}$, while the measured ratio rises from $1.02$ to $1.25$. The per-band shape mss, unlike their
difference, also carry the long-wave tilt of the sea state. At the reference sea state
KuKa-Mod is up to $10\%$ high in both bands at $4$--$8\,\mathrm{m\,s^{-1}}$, within
$4\%$ at $8$--$16\,\mathrm{m\,s^{-1}}$, and $4$--$8\%$ low at
$16$--$20\,\mathrm{m\,s^{-1}}$ (Fig.~\ref{fig:dualfreq}b). The matched reductions are not stratified by $\Hs$. The high-wind shortfall
has the sign of a long-wave tilt, which raises the shape mss of both bands by nearly the same
amount and leaves the shape difference nearly unchanged (Section~\ref{sec:val-hs}). The
low-wind excess cannot come from that tilt, so the misfit is not attributed to sea
state alone.

\section{Evaluation and Validation}
\label{sec:val}

The fitted model is now evaluated along incidence, band and wind. The test with
the Ka data withheld is the out-of-sample result and comes first. The transferred
Ka shape is an internal-consistency check, since the matched shape difference
enters the joint fit.

\subsection{What one band determines}
\label{sec:val-holdout}
\begin{figure*}[t]
  \centering
  \includegraphics[width=\textwidth]{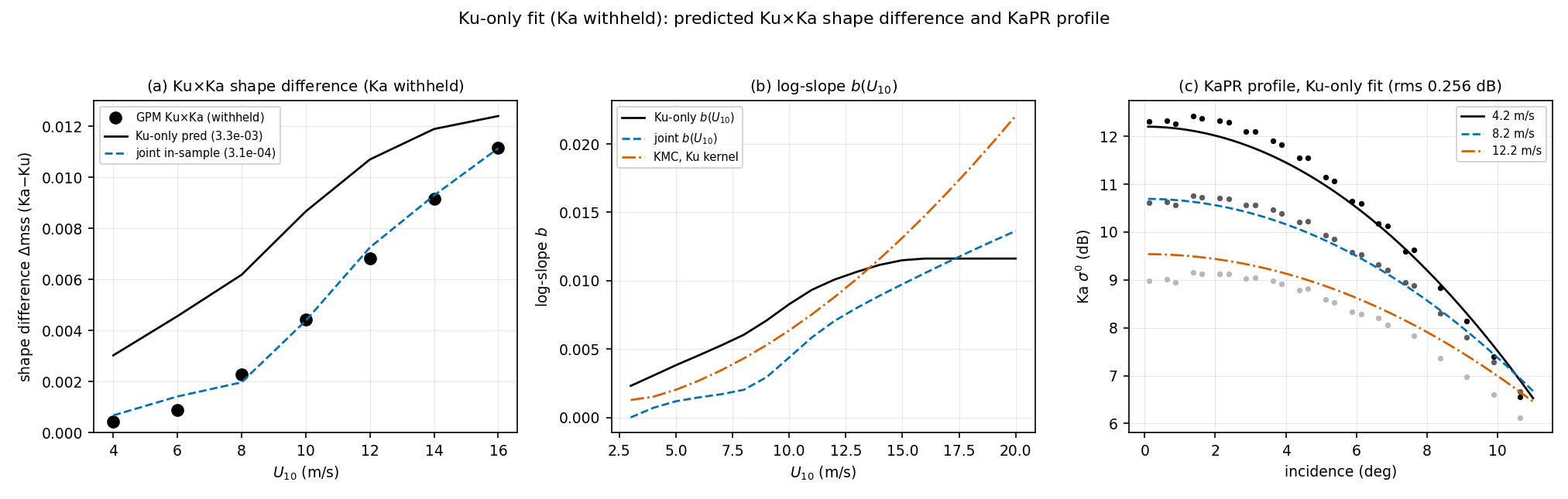}
  \caption{Ku$\to$Ka test with the Ka data withheld. (a)~The shape difference: the
  Ku-only prediction overshoots at low wind and meets the data only at high wind,
  whereas the joint fit matches from $8\,\mathrm{m\,s^{-1}}$ upward. (b)~The
  log-slope $b(\Uten)$: the Ku-only $b$ is inflated at low wind relative to the
  joint value and the KMC log-slope at the Ku kernel. (c)~The KaPR profile ($0$--$11^\circ$) with one
  fitted Ka level and no shape freedom: shape RMSD $0.26\dB$ over $611$ bins.}
  \label{fig:holdout}
\end{figure*}

To isolate what a single band constrains, the model is fitted to the Ku profile
only and $a(\Uten)$, $b(\Uten)$ are frozen. The frozen parameters then predict the withheld matched shape difference, with no Ka
calibration. The KaPR profile shape is also evaluated after one constant Ka level
is fitted (Table~\ref{tab:holdout}, Fig.~\ref{fig:holdout}). The Ku-only fit reproduces the
Ku profile as well as the joint fit ($0.105\dB$). Its predicted shape difference,
however, is far too large at low wind, $+626\%$ at $4\,\mathrm{m\,s^{-1}}$, falling
to $+11\%$ at $16\,\mathrm{m\,s^{-1}}$. A second Ku-only fit, started from
other initial nodes (Part~I), gives $+798\%$ and $+10\%$, so the low-wind failure
does not depend on the start, although its size does. This is the weak one-band constraint anticipated in
Part~I: the Ku profile leaves the split between coefficient and log-slope poorly
determined at low wind, and the one-band fit absorbs the profile shape into an
inflated $b$. Only at high wind, where the log-slope dominates the shape, do the
two fits converge. The KaPR profile shape is less sensitive. With one Ka level
fitted, its shape RMSD over $0$--$11^\circ$ is $0.26\dB$ against $0.21\dB$ for the
joint fit ($611$ bins). A near-nadir single-band profile therefore does not
determine the Ku$\to$Ka transfer at low-to-moderate wind, and KuKa-Mod is a joint
Ku--Ka retrieval rather than a one-band extrapolation.

\begin{table}[t]
  \centering
  \caption{Ku$\to$Ka test with the Ka data withheld. The Ku-only fit reproduces
  the Ku profile as well as the joint fit but fails to predict the shape difference
  at low wind.}
  \label{tab:holdout}
  \footnotesize
  \setlength{\tabcolsep}{3pt}
  \begin{tabular}{@{}p{0.56\columnwidth}cc@{}}
    \toprule
     & Ku-only & joint \\
     & (Ka withheld) & (Ka in-sample) \\
    \midrule
    Ku profile wRMSD (dB)                & 0.105 & 0.105 \\
    $\dmss$ RMSD                         & $3.3\times10^{-3}$ & $3.1\times10^{-4}$ \\
    \;$\dmss$ error at $4\,\mathrm{m\,s^{-1}}$           & $+626\%$ & --- \\
    \;$\dmss$ error at $16\,\mathrm{m\,s^{-1}}$          & $+11\%$  & --- \\
    KaPR shape RMSD, $0$--$11^\circ$, one Ka level (dB) & 0.26 & 0.21 \\
    \bottomrule
  \end{tabular}
\end{table}

\subsection{Incidence and wind residuals}
\label{sec:val-inc}
Over the fitted band the residuals are small. The Ku profiles hold an RMSD of bin
means of $0.09\mbox{--}0.12\dB$ over the specular core and $0.07\mbox{--}0.13\dB$ on the
$11$--$18^\circ$ flank (Fig.~\ref{fig:compare}, Table~\ref{tab:compare}). At Ku,
GO with $\mss=s^2(\rstar)$ agrees with the full integral to within $0.10\dB$ over
$3$--$8\,\mathrm{m\,s^{-1}}$ and $0$--$10^\circ$. The difference reaches $0.28\dB$ at
$12\,\mathrm{m\,s^{-1}}$ and $11^\circ$.

The residuals also depend on the footprint. A radar cell averages $\snot$ over a
heterogeneous wind of mean $\bar\Uten$ and standard deviation $\sigma_U$. To second
order in $\sigma_U$ this average renormalizes the structure function (Part~I),
\begin{equation}
  \Dtwo^{\rm eff}(r)=\Dtwo(r;\bar\Uten)
   +\tfrac12\,\partial_U^2\Dtwo\,\sigma_U^2
   -\tfrac14\,\qz^2\,(\partial_U\Dtwo)^2\,\sigma_U^2,
  \label{eq:d2eff}
\end{equation}
where $\partial_U$ is the derivative with respect to wind. With the gustiness law
of Part~I, which scales as $\ell^{1/3}$ with the footprint size $\ell$, the
correction is evaluated by Gaussian wind averaging, has no fitted amplitude and
stays below $0.05\dB$. Its last term is the image in $\Dtwo^{\rm eff}$ of the wind-mixture kurtosis, present even
for Gaussian increments, and the error of a reference wind produces the same mixing
among collocations, as \citet{coxmunk1954,coxmunk1956} already noted for their slope peakedness. Applied as a
differential from the GPM footprint, it lowers the low-wind incidence residual of
calibrated SWIM by about $8\%$. Like the wind-variability term of scatterometry
\citep{portabella2006}, it is prescribed rather than fitted.

\subsection{Wind reference and air--sea stability}
\label{sec:val-stab}
The released KuKa-Mod is referenced to the ERA5 $10$-m wind $\Uten$, whereas scatterometer GMFs are referenced to the
neutral-equivalent or stress-equivalent wind \citep{dekloe2017}, which follows the
surface stress. At a given $\Uten$ the stress is higher in unstable air, where the
sea is warmer than the air, so the surface is rougher than $\Uten$ suggests. Inside
the hinge a rougher surface is darker, so if near-nadir $\snot$ follows the stress,
the GMF residual at fixed $\Uten$ should fall as the sea-surface temperature (SST) minus the
$2$-m air temperature, $\mathrm{SST}-T_{2\rm m}$, rises.

This test was made on $15\%$ subsamples of KuPR 2021--2022 ($2.0\times10^6$) and of
calibrated SWIM 2021 ($6.8\times10^6$). The neutral-equivalent wind $U_{10N}$ was
computed from the ERA5 wind, sea-surface temperature, $2$-m air temperature,
dewpoint and pressure with a simplified Monin--Obukhov iteration after the Coupled Ocean--Atmosphere Response
Experiment (COARE) algorithm \citep{fairall2003}. The GMF is read at $U_{10N}$ through a relabelling of the
wind axis by the conditional mean of $\Uten$ given $U_{10N}$, measured on KuPR. Against
$\Uten$, the residual darkens in unstable air by $0.044\dB\,\mathrm{K^{-1}}$ at
$2^\circ$, and the slope decreases to near zero at $10^\circ$ in both sensors
(Table~\ref{tab:stab}), and reaches $-0.10\dB\,\mathrm{K^{-1}}$ at
$2$--$4\,\mathrm{m\,s^{-1}}$. With
$U_{10N}$ as the wind, the slope decreases by $77$--$90\%$ at $2$--$8^\circ$, and the
small $10^\circ$ slope changes sign. This supports a stress-referenced near-nadir
$\snot$, as for scatterometers, although both winds derive from ERA5.

\begin{table}[t]
  \centering
  \caption{Stability dependence of the GMF residual: slope of
  $\snot-\snot_{\rm GMF}$ against $\mathrm{SST}-T_{2\rm m}$ at fixed wind
  (dB\,K$^{-1}$; count-weighted mean of the $6$--$8$, $8$--$10$ and
  $10$--$13\,\mathrm{m\,s^{-1}}$ bins), with the GMF read at the ERA5 wind $\Uten$ or
  at the neutral-equivalent wind $U_{10N}$.}
  \label{tab:stab}
  \footnotesize
  \setlength{\tabcolsep}{4pt}
  \begin{tabular}{@{}lcccc@{}}
    \toprule
    & \multicolumn{2}{c}{KuPR} & \multicolumn{2}{c}{SWIM} \\
    Incidence & $\Uten$ & $U_{10N}$ & $\Uten$ & $U_{10N}$ \\
    \midrule
    $2^\circ$  & $-0.044$ & $-0.007$ & $-0.044$ & $-0.005$ \\
    $4^\circ$  & $-0.043$ & $-0.009$ & $-0.039$ & $-0.004$ \\
    $6^\circ$  & $-0.033$ & $-0.007$ & $-0.031$ & $-0.005$ \\
    $8^\circ$  & $-0.022$ & $-0.005$ & $-0.019$ & $-0.002$ \\
    $10^\circ$ & $-0.006$ & $+0.001$ & $-0.005$ & $+0.002$ \\
    \bottomrule
  \end{tabular}
\end{table}

In the KuPR sample $U_{10N}-\Uten$ averages $+0.25\,\mathrm{m\,s^{-1}}$, and the
offset is geographically organized. A retrieval that combines KuKa-Mod with a
scatterometer GMF should therefore use a common reference. The scatterometer
stress-equivalent wind $U_{10S}=U_{10N}\sqrt{\rho_a/\rho_0}$, with
$\rho_0=1.225\,\mathrm{kg\,m^{-3}}$ \citep{dekloe2017}, provides it. In the
$1.69\times10^6$ KuPR samples of 2021--2022 with a scatterometer wind, the
scatterometer wind lies on average $0.07\,\mathrm{m\,s^{-1}}$ below the ERA5
estimate of $U_{10S}$.

The Ku core was refitted against $U_{10S}$. The difference between the profiles
binned in $U_{10S}$ and in $\Uten$ on the collocated samples, smoothed, is added to
the full ERA5-binned profile, and the $\dmss$ constraint is placed at the mean
$U_{10S}$ of each ERA5 wind class. The nonzero $b$ nodes change by at most $4\%$, and
the $a$ nodes by at most $4\%$ except at $16\,\mathrm{m\,s^{-1}}$ ($-9\%$). The
profile wRMSD falls from $0.105$ to $0.103\dB$, while the $\dmss$ RMSD rises from
$3.1$ to $3.5\times10^{-4}$. Within $10^\circ$ the Ku $\snot$ changes by at most
$0.03\dB$ at $7$--$16\,\mathrm{m\,s^{-1}}$, less than the profile residual, and by up
to $0.15\dB$ at $3$ and $20\,\mathrm{m\,s^{-1}}$. The released parameters, figures and
tables keep the ERA5 reference. A Ku table of the refitted core is released for
retrieval with scatterometer winds and supersedes the earlier $U_{10N}$ table. The
directional module was not refitted, and the Ka level is retained after a test against
$U_{10S}$ (Section~\ref{sec:val-band}).

\subsection{Band transfer to Ka}
\label{sec:val-band}
\begin{figure*}[t]
  \centering
  \includegraphics[width=\textwidth]{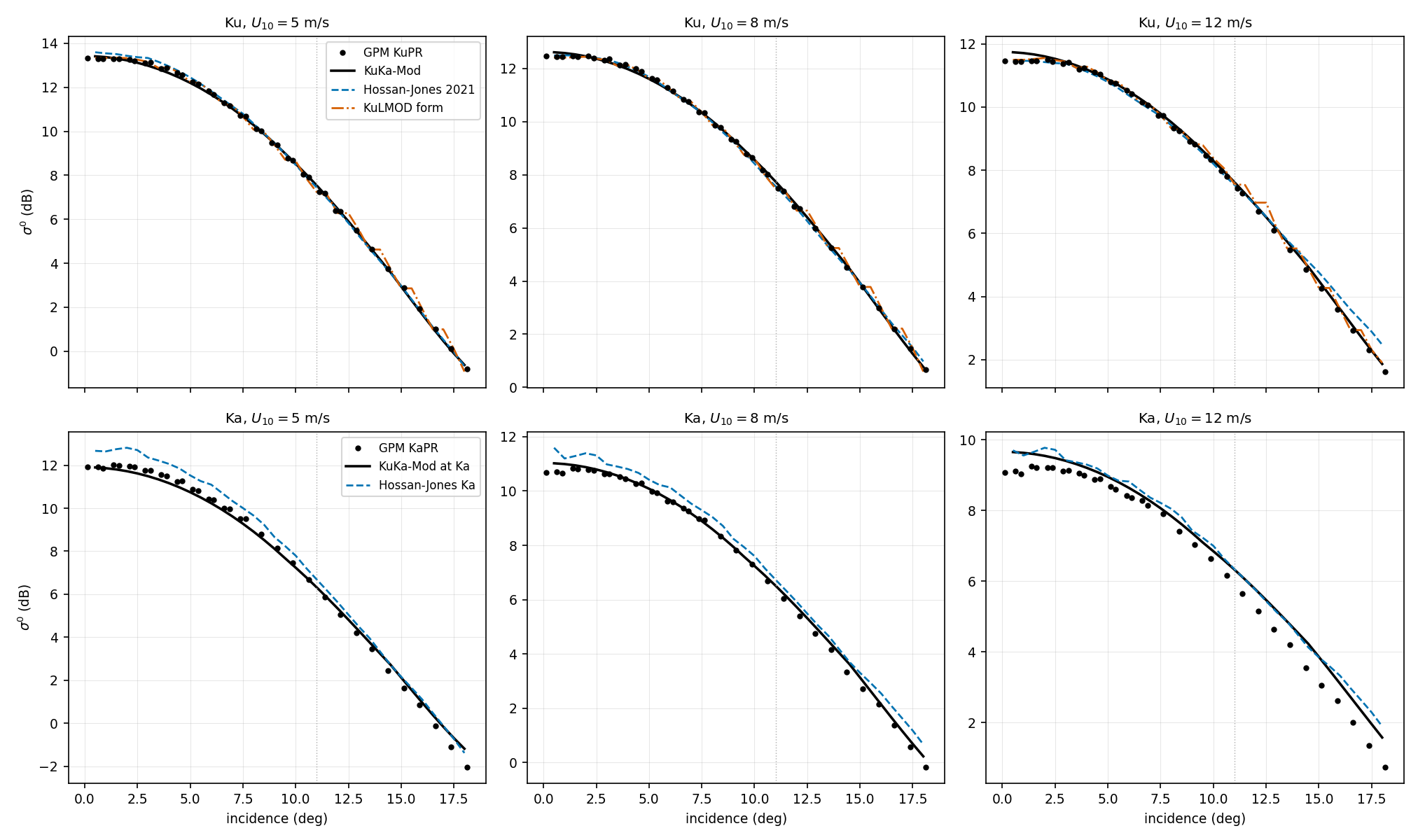}
  \caption{KuKa-Mod against GPM data and published GMFs over $0$--$18^\circ$
  (dotted: the $11^\circ$ boundary between specular core and flank). Top, Ku (the
  fitted band): KuPR (markers), KuKa-Mod, the re-referenced \citet{hossan2021}
  model and a refitted KuLMOD power law at three winds. Bottom, Ka (transferred
  with one fitted level): KaPR (markers), KuKa-Mod with no shape refit, and the
  \citet{hossan2021} Ka model. RMSD values are given in Table~\ref{tab:compare}.}
  \label{fig:compare}
\end{figure*}

The model is referenced at Ku and transferred to Ka with no band-specific shape
parameter. The same $\Dtwo(r)$ is read at $\rstar_{\rm Ka}$ ($1.0$--$0.5$\,cm)
instead of $\rstar_{\rm Ku}$ ($2.5$--$1.4$\,cm). With one Ka level
$R_{\rm eff,Ka}^2\simeq0.43$ and no shape refit, the model tracks KaPR to
$0.13\mbox{--}0.32\dB$ over the specular core (Fig.~\ref{fig:compare},
Table~\ref{tab:compare}). On the Ka flank beyond $11^\circ$ the residual grows to
$0.64\dB$ at $12\,\mathrm{m\,s^{-1}}$, almost entirely a positive bias, because
the Bragg ramp fitted on Ku transfers least well.

Above $16\,\mathrm{m\,s^{-1}}$ the matched data are not fitted, yet KuKa-Mod follows them within $4\%$ up to
$21\,\mathrm{m\,s^{-1}}$ (Fig.~\ref{fig:dualfreq}a).

\begin{figure}[t]
  \centering
  \includegraphics[width=\columnwidth]{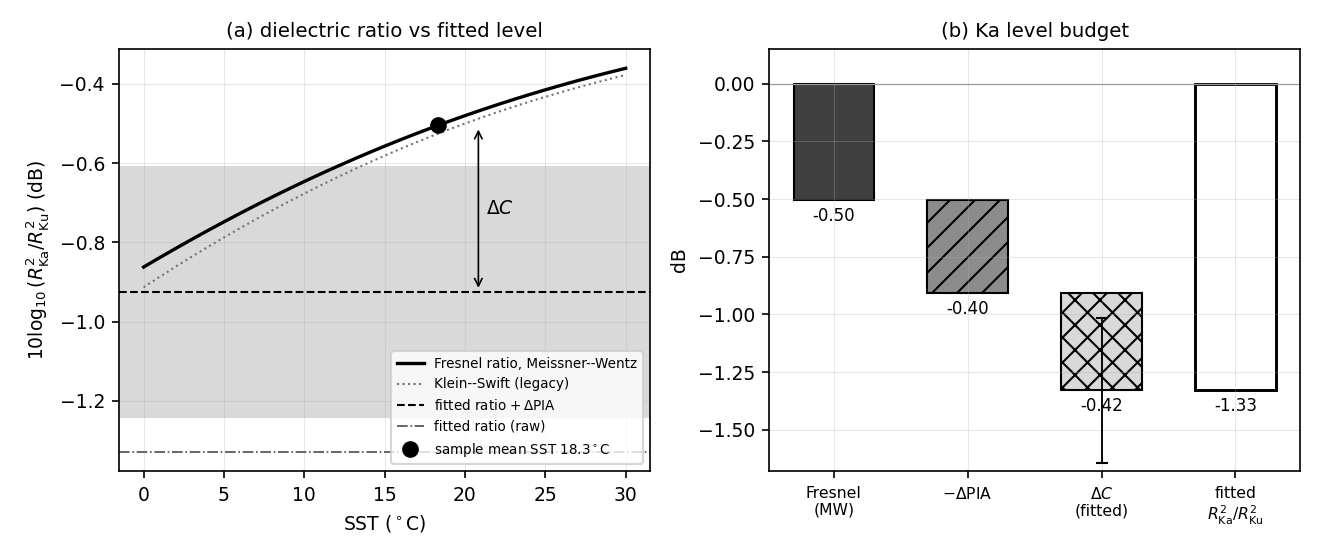}
  \caption{Ka level budget on the matched data. (a)~Ka/Ku nadir reflectivity ratio
  against sea-surface temperature from the \citet{meissner2004} dielectric (solid;
  Klein--Swift \citep{klein1977} dotted), the fitted ratio corrected for the differential clear-sky
  PIA (dashed, $\pm0.31\dB$ band) and the raw fitted ratio (dash-dot); marker at
  the mean $18.3^\circ$C. (b)~Budget: Fresnel ratio $-0.505\dB$, differential
  attenuation $-0.403\dB$ and residual calibration term $\Delta C=-0.42\pm0.31\dB$
  sum to the fitted $-1.33\dB$.}
  \label{fig:kalevel}
\end{figure}

The only band-specific quantity left is the Ka level, and an independent budget
explains most of it (Fig.~\ref{fig:kalevel}). The fitted level ratio is
$10\log_{10}(R^2_{\rm eff,Ka}/R^2_{\rm eff,Ku})=-1.33\dB$. The seawater dielectric
\citep{meissner2004} gives a Ka/Ku reflectivity ratio of $-0.50\dB$ at the mean
sea-surface temperature of the matched data ($18.3^\circ$C). The clear-sky two-way
PIA, measured from the non-precipitation correction of $3.7\times10^8$ matched
pairs, is $0.16\dB$ at Ku and $0.56\dB$ at Ka, and varies by less than $0.01\dB$
across $0$--$10^\circ$. The dielectric and the differential attenuation thus
explain $-0.91$ of the $-1.33\dB$. The remaining
$\Delta C=-0.42\pm0.31\dB$ is of the order of the absolute calibration
uncertainty of the two radars, but the Ka level matched per wind drifts from
$-1.10\dB$ at $5\,\mathrm{m\,s^{-1}}$ to $-1.56\dB$ at
$12\,\mathrm{m\,s^{-1}}$, which no calibration constant can do. This residual sets the Ka core RMSD of Table~\ref{tab:compare}. It amounts to a
transferred Ka wind sensitivity about $20\%$ too weak between $5$ and
$12\,\mathrm{m\,s^{-1}}$, a wind bias of about $1.3\,\mathrm{m\,s^{-1}}$ at
$12\,\mathrm{m\,s^{-1}}$ for a Ka retrieval with the released level.

The GPM--scatterometer collocations (Section~\ref{sec:data-models}) extend this test
to $25\,\mathrm{m\,s^{-1}}$ with an independent wind. In $9.5\times10^7$ paired
footprints, the measured Ka--Ku difference
over $1$--$9^\circ$ falls from $-1.53\dB$ at $8\,\mathrm{m\,s^{-1}}$ to $-2.36\dB$ at
$20\,\mathrm{m\,s^{-1}}$ and $-2.53\dB$ at $25\,\mathrm{m\,s^{-1}}$. Read at $U_{10S}$, the
data-minus-model difference stays between $-0.12$ and $-0.15\dB$ from $10$ to
$20\,\mathrm{m\,s^{-1}}$ and reaches $-0.27\dB$ at $25\,\mathrm{m\,s^{-1}}$. Below
$10\,\mathrm{m\,s^{-1}}$, where the stratified sample is thin, it changes sign, which is
the weak Ka wind sensitivity again. With the ERA5 wind it is $-0.18$ to $-0.21\dB$ over
$15$--$20\,\mathrm{m\,s^{-1}}$. The Ka--Ku $\snot$ difference thus holds to within $0.3\dB$ up
to $25\,\mathrm{m\,s^{-1}}$ with either reference, although the misfit grows above
$20\,\mathrm{m\,s^{-1}}$. The change of reference does not remove the Ka residual: the
Ka level matched over $2$--$9^\circ$ moves by $0.03\dB$ from ERA5 to $U_{10S}$, and
the level matched per wind still falls from $+0.34$ to $-0.19\dB$ between $5$ and
$12\,\mathrm{m\,s^{-1}}$, against $+0.08$ to $-0.05\dB$ at Ku. The released Ka level is
kept. The stress-equivalent wind improves the core only at high wind: at
$16\,\mathrm{m\,s^{-1}}$, with each reference at its own matched Ka level, the
model-minus-data bias over $0$--$11^\circ$ falls from
$+0.75$ to $+0.52\dB$ at Ka and from $+0.28$ to $+0.21\dB$ at Ku, while at
$5$--$12\,\mathrm{m\,s^{-1}}$ the two references agree within $0.02\dB$.

\begin{table}[t]
  \centering
  \caption{RMSD of KuKa-Mod minus GPM bin means (dB) by incidence range and wind,
  with the bias (model minus data) in parentheses and the number $N$ of
  $0.25^\circ$ bins. Ku holds $\lesssim0.13\dB$ at the tabulated winds. The Ka core residual
  is dominated by a wind-dependent level error, and the Ka flank at
  $12\,\mathrm{m\,s^{-1}}$ by a $+0.63\dB$ bias.}
  \label{tab:compare}
  \footnotesize
  \setlength{\tabcolsep}{3pt}
  \begin{tabular}{@{}lccc@{}}
    \toprule
    & $\Uten=5$ & $8$ & $12\,\mathrm{m\,s^{-1}}$ \\
    \midrule
    Ku, $0$--$11^\circ$ ($N{=}30$)  & 0.10 ($-0.02$) & 0.09 ($0.00$) & 0.12 ($+0.08$) \\
    Ku, $11$--$18^\circ$ ($N{=}10$--$11$) & 0.09 ($-0.02$) & 0.07 ($+0.02$) & 0.13 ($+0.13$) \\
    Ka, $0$--$11^\circ$ ($N{=}24$--$25$)  & 0.21 ($-0.19$) & 0.13 ($+0.06$) & 0.32 ($+0.29$) \\
    Ka, $11$--$18^\circ$ ($N{=}10$) & 0.38 ($+0.33$) & 0.25 ($+0.24$) & 0.64 ($+0.63$) \\
    \bottomrule
  \end{tabular}
\end{table}

\subsection{High-wind level and instrument calibration}
\label{sec:val-highwind}
Above $16\,\mathrm{m\,s^{-1}}$ KuKa-Mod extrapolates the logarithmic structure function.
Against scatterometer winds the extrapolated level is too bright, and since $\snot$
decreases with wind, a retrieval based on it reads high. For operational use the
azimuth shape of KuKa-Mod is kept and its level is replaced by a measured one,
\begin{equation}
  \snot_{\rm op}(U_{10S},\Phi,\theta)=\snot_{\rm KuKa}(U_{10S},\Phi,\theta)\,10^{\Delta_{\rm hw}(U_{10S},\theta)/10},
  \label{eq:highwind}
\end{equation}
where $\snot_{\rm KuKa}$ is the model at the reference sea state, read at $U_{10S}$
with the Ku core of Section~\ref{sec:val-stab} and the released Ka level, and
$\Delta_{\rm hw}$ is a smooth level correction in decibels. $\Delta_{\rm hw}$ is zero
below $4\,\mathrm{m\,s^{-1}}$, is blended in between $4$ and $8\,\mathrm{m\,s^{-1}}$,
and applies in full above.

Conditional means of $\snot$ at a given reference wind are biased in the sparse
tail: the random error of the reference dilutes them, the true wind being lower on
average at high reference wind, and the table comes out too bright. Where possible,
$\Delta_{\rm hw}$ is therefore obtained by matching quantiles of $\snot$ to quantiles of
$U_{10S}$ in each $0.25^\circ$ incidence bin \citep{freilich1994}, after rescaling the
wind quantiles for the different noise of the two variables. The error variances
come from triple collocation of $\snot$, $U_{10S}$ and ERA5 \citep{stoffelen1998};
the $U_{10S}$ error is $0.41$--$0.47\,\mathrm{m\,s^{-1}}$ at Ku and mostly
$0.36$--$0.43\,\mathrm{m\,s^{-1}}$ at Ka. The scatterometer winds are shifted to the
radar time with ERA5, which couples the wind errors; this covariance is not quantified,
variance rescaling need not correct skewed quantiles, and the resulting uncertainty of
$\Delta_{\rm hw}$ is not propagated. Beyond $6^\circ$ at Ku and $7.5^\circ$ at Ka a single
footprint is too noisy for the collocation, its equivalent wind error exceeding
$2.5\,\mathrm{m\,s^{-1}}$; further out the correlation with the wind weakens, and
beyond about $10.5^\circ$ $\snot$ is wind-blind. There $\Delta_{\rm hw}$ is taken from weighted medians of
$\snot$ in $0.5\,\mathrm{m\,s^{-1}}$ bins of the reference wind, which keep part of
the dilution. Details are given in Appendix~\ref{app:proc}.

\begin{figure}[t]
  \centering
  \includegraphics[width=\columnwidth]{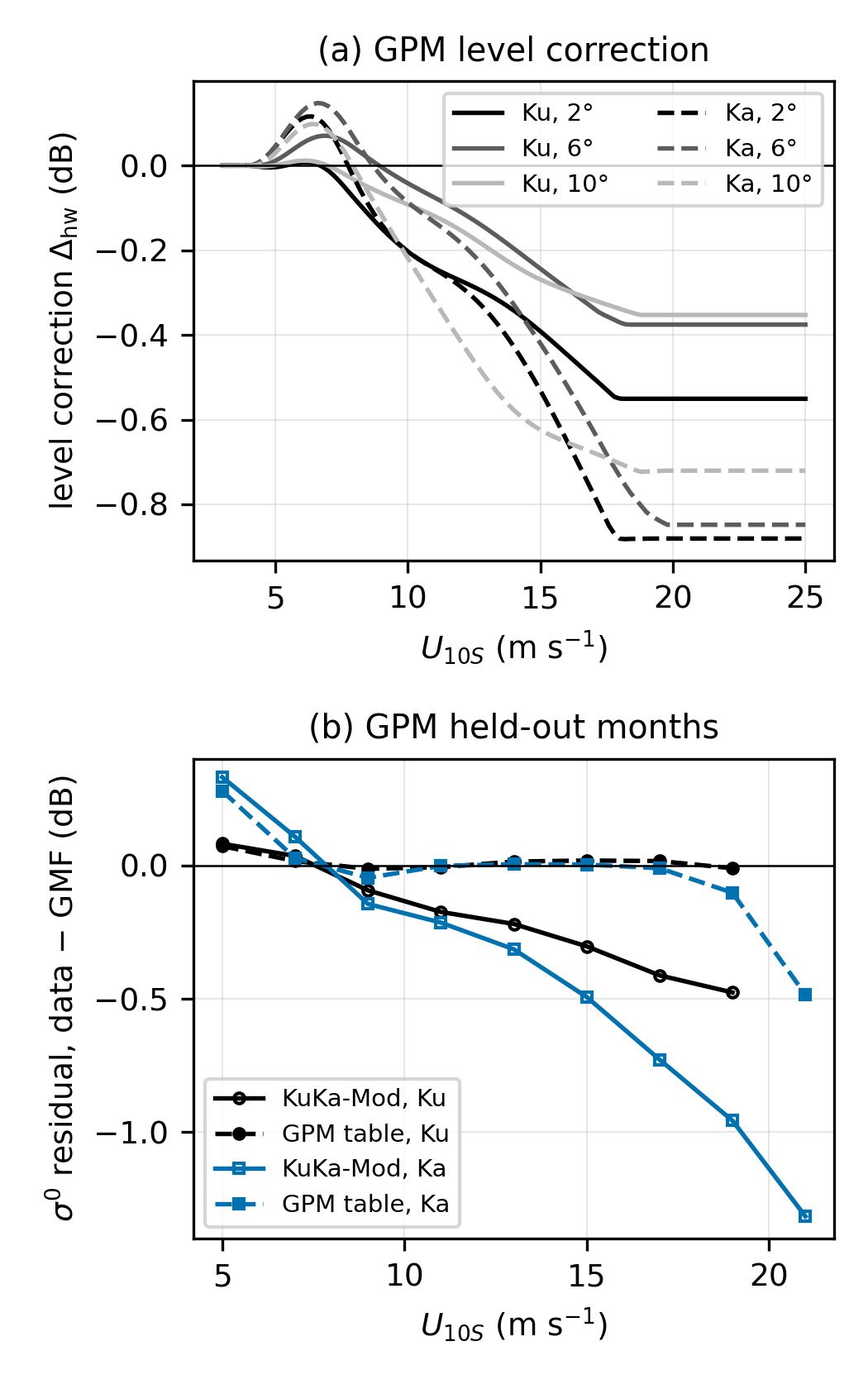}
  \caption{High-wind level of KuKa-Mod on GPM against scatterometer $U_{10S}$.
  (a)~Level correction $\Delta_{\rm hw}$ of \eqref{eq:highwind} at $2$, $6$ and
  $10^\circ$ (grey level), Ku (solid) and Ka (dashed); it is held constant beyond
  the data support. (b)~Mean $\snot$ residual on the withheld months, data minus GMF, against
  the reference wind, in the quantile-matched incidence bins: KuKa-Mod (solid, open) and the
  GPM tables (dashed, filled), Ku (circles) and Ka (squares).}
  \label{fig:highwind}
\end{figure}

The correction is measured on the GPM sample of Section~\ref{sec:data-models}, the data
KuKa-Mod was fitted to: $1.02\times10^8$ Ku and $1.08\times10^8$ Ka footprints, of
which $1.5\times10^6$ and $1.7\times10^6$ have $U_{10S}\ge20\,\mathrm{m\,s^{-1}}$. Odd
months are fitted and even months are withheld. On the withheld months, averaged over the
matched quantiles of the quantile-matched incidence bins, the mean $\snot$ residual (data minus KuKa-Mod) above $12\,\mathrm{m\,s^{-1}}$ is $-0.29\dB$ at
Ku and $-0.47\dB$ at Ka, the equivalent wind offsets being $1.4$ and
$2.1\,\mathrm{m\,s^{-1}}$ high, and
$+0.02$ and $0.00\dB$ with $\Delta_{\rm hw}$ (Fig.~\ref{fig:highwind}b). From $8$ to
$20\,\mathrm{m\,s^{-1}}$ the residual with $\Delta_{\rm hw}$ stays within $0.02\dB$ at
Ku and $0.1\dB$ at Ka. In the sparse $20$--$22\,\mathrm{m\,s^{-1}}$ bin the Ka table
is still $0.48\dB$ too bright ($3.5\,\mathrm{m\,s^{-1}}$ wind equivalent), and the Ku
assessment ends at $20\,\mathrm{m\,s^{-1}}$. This tests the distributional calibration
in the quantile-matched bins (about $\theta\lesssim6^\circ$ at Ku and $7.5^\circ$ at Ka),
excluding the median-fallback bins; it does not measure retrieval accuracy. Below $8\,\mathrm{m\,s^{-1}}$, where $\Delta_{\rm hw}$ is tapered and at Ka slightly
positive, up to about $+0.15\dB$ near $6\,\mathrm{m\,s^{-1}}$, the Ka residual at
$4$--$6\,\mathrm{m\,s^{-1}}$ is $+0.28\dB$, against $+0.33\dB$ for KuKa-Mod. The data support ends near $18$--$20\,\mathrm{m\,s^{-1}}$, and
$\Delta_{\rm hw}$ is held constant above it (Fig.~\ref{fig:highwind}a): at $2$, $6$ and
$10^\circ$ the plateau is $-0.55$, $-0.38$ and $-0.35\dB$ at Ku and $-0.88$, $-0.85$
and $-0.72\dB$ at Ka, and it reaches $-0.72\dB$ at Ku and $-1.18\dB$ at Ka at
$0.5^\circ$. The larger Ka correction continues the weak Ka wind sensitivity of
Section~\ref{sec:val-band}. These two tables are the reference high-wind level of KuKa-Mod at
Ku and Ka.

A given instrument does not reproduce the GPM level exactly: its gain, processing and
footprint differ from those of the DPR, and a calibration to KuPR at moderate winds
(Appendix~\ref{app:proc} for SWIM) need not hold in the tail. For another near-nadir
instrument we recommend starting from the GPM table of its band and fitting an
instrument term against collocated scatterometer $U_{10S}$ with the same estimator.
The table obtained in this way for SWIM is released with the model. $\Delta_{\rm hw}$
is a calibration, not part of the physical model, and the KuKa-Mod parameters are
unchanged.

\section{Sea State and Azimuth}
\label{sec:axes}

Beyond wind, incidence and band, two further variables modulate the near-nadir
cross-section: the sea state and the wind-relative azimuth. Both are evaluated on the sensor that constrains them best:
SWIM for its dense azimuth sampling and collocated $\Hs$, and GPM for the level
and the band ratio.

\subsection{Sea state}
\label{sec:val-hs}
This subsection measures how the near-nadir cross-section responds to the
significant wave height $\Hs$ at fixed wind, and tests what the model predicts about
that response without a fitted sea-state coefficient. The long waves tilt the
short-wave facets, and over the footprint their slopes add to the short-wave slopes
\citep{chu2012}.
In the model this tilt shifts the coefficient $a$ of the structure function $\Dtwo(r)$
away from its value at the reference sea state $\Hs\in[1.5,3.5]$\,m,
\begin{equation}
  a(\Uten,\Hs)=a_{\rm ref}(\Uten)+\Delta\sigma^2_{\rm long}(\Uten,\Hs),
  \label{eq:hsmss}
\end{equation}
where $\Delta\sigma^2_{\rm long}$ is the long-wave slope variance in excess of its
mean over the reference window, positive at higher sea states.
For $b\ll\msseff$ it adds nearly the same amount to $\msseff$ in both bands, the additive
form of \citet{tran2007}, following \citet{tranchapron2006}. The geometric-optics
core $\snot\propto\msseff^{-1}\exp(-\tan^2\theta/\msseff)$ then gives, in dB
($\snot_{\rm dB}=10\log_{10}\snot$),
\begin{equation}
  \frac{\partial\snot_{\rm dB}}{\partial\Hs}
   =\frac{10}{\ln10}\,\frac{\tan^2\theta-\msseff}{\msseff^{2}}\,
    \frac{\partial\Delta\sigma^2_{\rm long}}{\partial\Hs}.
  \label{eq:hsresp}
\end{equation}
For a positive coefficient $\partial\Delta\sigma^2_{\rm long}/\partial\Hs$, the
coefficient sets the size of the response but not its sign. To first order in the anomaly, two properties therefore follow from the wind core
alone: the response changes sign at the hinge $\tan^2\theta=\msseff$ of the reference
sea state, and the ratio of the Ka to the Ku response in dB is fixed by the two bands'
$\msseff$. We test both, and leave the
coefficient itself unfitted for the reasons given at the end of this subsection.

\begin{figure*}[t]
  \centering
  \includegraphics[width=0.8\textwidth]{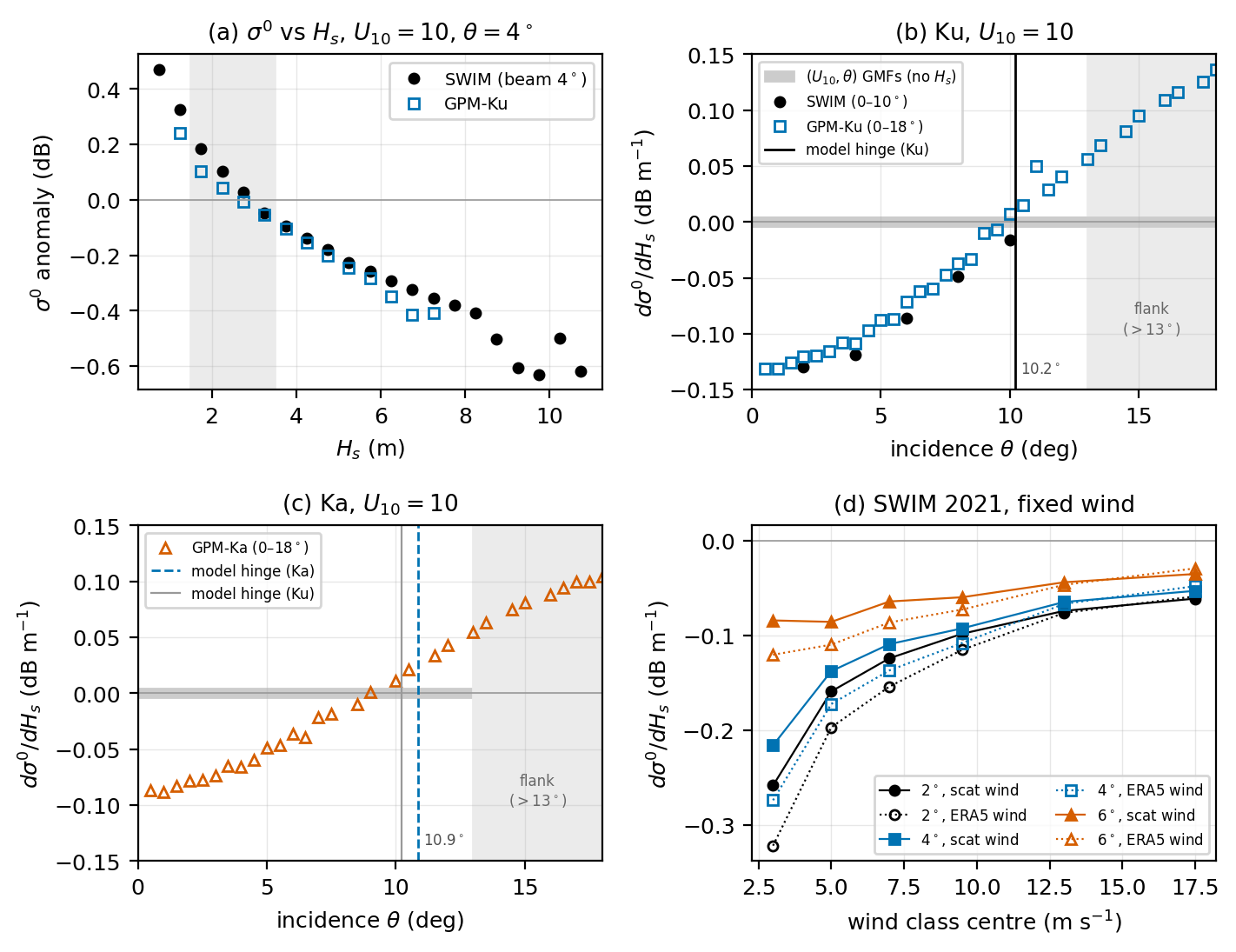}
  \caption{Measured sea-state response of the near-nadir cross-section and the
  model zero crossings, which need no sea-state coefficient \eqref{eq:hsresp}.
  (a)~$\snot$ anomaly against $\Hs$ at $10\,\mathrm{m\,s^{-1}}$ and $4^\circ$: SWIM
  (filled) and GPM-KuPR (open), about the reference window (shaded).
  (b)~$d\snot/d\Hs$ against incidence at $10\,\mathrm{m\,s^{-1}}$ (count-weighted
  regressions over $\Hs=1$--$4$\,m): SWIM and GPM-Ku, with the Ku zero crossing of the full
  model ($10.2^\circ$, vertical line). GMFs without an $\Hs$ term lie at zero (grey).
  Beyond $13^\circ$ (shaded) lies the empirical flank, where the simple tilt
  interpretation is less reliable.
  (c)~The same at Ka: KaPR with the Ka model zero crossing ($10.9^\circ$, dashed) and the Ku
  one (grey); the observed response turns positive near $8.9^\circ$.
  (d)~SWIM 2021 $d\snot/d\Hs$ at $2$, $4$ and $6^\circ$ per wind class, at fixed
  scatterometer wind (filled, solid) and at fixed ERA5 wind (open, dotted). Wind
  speeds in (a)--(c) are ERA5.}
  \label{fig:seastate}
\end{figure*}

As shown in Fig.~\ref{fig:seastate}a, at $10\,\mathrm{m\,s^{-1}}$ and $4^\circ$
SWIM and GPM-Ku agree on the decrease of $\snot$ with $\Hs$ up to about $6$\,m.
Over $\Hs=1$--$4$\,m the $2^\circ$ slope is $-0.13\dB\,\mathrm{m^{-1}}$ for SWIM and
$-0.12\dB\,\mathrm{m^{-1}}$ for GPM-Ku (Fig.~\ref{fig:seastate}b). The response
weakens with incidence and changes sign at $9.7^\circ$ for GPM-Ku. The geometric-optics hinge lies at $10.8^\circ$ (Ku) and $11.4^\circ$ (Ka) at
$10\,\mathrm{m\,s^{-1}}$. With the full logarithmic Kirchhoff integral instead of its GO limit, a
small level increment changes the sign of the response at $10.2^\circ$ (Ku) and
$10.9^\circ$ (Ka). The full model thus places the Ku sign change within about
$0.5^\circ$ of the measured one, with no sea-state parameter. This sign change is the
reversal of the $\snot$--$\Hs$ response documented by \citet{tran2007}.
At Ka the prediction fails. A band-common tilt gives a Ka/Ku response ratio of $0.91$
over $2$--$6^\circ$, whereas GPM measures $0.59$, and the KaPR response changes sign
near $8.9^\circ$, inside the Ku one rather than outside it
(Fig.~\ref{fig:seastate}c). KaPR $\snot$ is not corrected for path attenuation, and
attenuation may grow with sea state. Part of the Ka response may therefore not be
carried by a band-common slope variance. The $(\Uten,\theta)$ GMFs of
\citet{hossan2021} and \citet{ren2021} omit this response of the inner beams.

The size of the response depends on the reference wind. Triple collocation of SWIM,
ERA5 and the KNMI scatterometer winds within $30$\,min in 2021 \citep{stoffelen1998}
gives random errors of $0.86\,\mathrm{m\,s^{-1}}$ for ERA5 and $0.42\,\mathrm{m\,s^{-1}}$ for the
closest ASCAT, HY-2 or ScatSat pass. With the scatterometer wind held fixed, the $2^\circ$
slope falls from $-0.26\dB\,\mathrm{m^{-1}}$ at $2$--$4\,\mathrm{m\,s^{-1}}$ to $-0.06$
at $15$--$20\,\mathrm{m\,s^{-1}}$ (Fig.~\ref{fig:seastate}d). With the ERA5 wind it is
about $25\%$ steeper below $8\,\mathrm{m\,s^{-1}}$, and above $11\,\mathrm{m\,s^{-1}}$
both agree. The measured size of the response therefore depends on the reference
wind, while its persistence at high wind does not. At fixed ERA5 wind, $\Hs$ explains $6$--$11\%$ of the
residual $\snot$ variance on the $2$ and $4^\circ$ beams below $11\,\mathrm{m\,s^{-1}}$,
$2$--$6\%$ at $6^\circ$ and less than $2\%$ at $8$--$10^\circ$, out of sample.

The response is consistent with the long-wave slope variance of \eqref{eq:hsmss}
rather than the wave height itself. In SWIM 2021, at fixed scatterometer
wind, $\snot$ was regressed on the slope variance of wavenumbers below
$3\,\mathrm{rad\,m^{-1}}$, taken from the SWIM spectrum up to
$0.2\,\mathrm{rad\,m^{-1}}$ and from WAVEWATCH~III beyond. On the $2$--$6^\circ$
beams the fitted sensitivity follows the geometric-optics derivative of
\eqref{eq:hsresp} in incidence and wind at $0.45$ of its amplitude, with a
systematic spread of $\pm0.07$ across cells and wind classes. The $10^\circ$ beam
shows the predicted sign change at low wind. With the slope variance in the regression, the remaining $\Hs$
coefficient lies between $-0.04$ and $+0.01\dB\,\mathrm{m^{-1}}$. Wave height is
thus only a proxy for the slope variance, to which swell adds little. At a given $\Hs$
the long-wave slope variance also depends on the development of the sea, as buoy
spectra show for C-band altimeter returns \citep{gourrion2002cband}. In the GPM
collocations, at fixed $U_{10S}$ and $\Hs$, shorter ERA5 peak periods accompany darker
returns below about $8^\circ$ in both bands, consistent with a larger slope variance
in \eqref{eq:hsresp}.

These results support the qualitative tilt mechanism, but its amplitude remains
unresolved: the $0.45\pm0.07$ efficiency, the Ka/Ku ratio and the Ka zero crossing
preclude a quantitative confirmation, their uncertainties are not propagated here, and
for the reasons of Section~\ref{sec:disc-limits} no operational law is fitted.

\subsection{Azimuth}
\label{sec:val-dir}
\begin{figure*}[t]
  \centering
  \includegraphics[width=\textwidth]{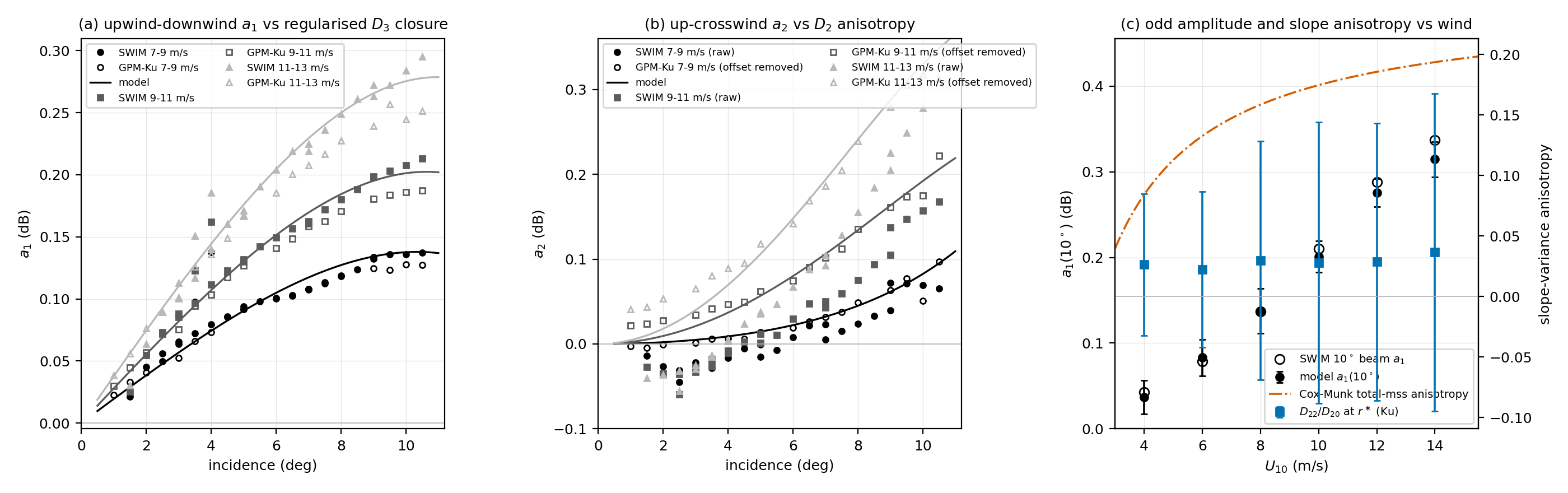}
  \caption{Azimuth harmonics of SWIM (filled, binned in incidence over the five beams) and GPM-Ku (open) for
  wind classes $7$--$9$, $9$--$11$ and $11$--$13\,\mathrm{m\,s^{-1}}$ (grey levels),
  against the model (lines). (a)~Upwind--downwind $a_1$ (dB), reproduced across all
  beams (RMSD $0.007$--$0.015\dB$). (b)~Up--crosswind $a_2$ (dB): the model,
  fitted to the offset-removed GPM-Ku $a_2$, reproduces the positive values and
  their wind growth; the small negative near-nadir SWIM $a_2$ is treated as a sensor offset
  (Fig.~\ref{fig:directional}). (c)~$a_1(10^\circ)$ and the anisotropy
  $D_{22}/D_{20}$ at the Ku kernel against wind (dash-dot: the Cox--Munk
  total-slope anisotropy, for scale).}
  \label{fig:azimuth}
\end{figure*}

The azimuth harmonics test the anisotropic terms of \eqref{eq:full_dir}. SWIM's
dense azimuth sampling fixes the form of the closure, while the amplitudes are set
on GPM-Ku, the radiometric reference. This split is possible because the odd
harmonic $a_1$ agrees between SWIM and GPM-Ku at every wind, whereas the even
harmonic $a_2$ carries a sensor-specific offset near nadir. With these settings the
odd $D_3$ reproduces $a_1(\theta)$ to $0.007$--$0.015\dB$ RMSD over the five beams,
including its rise through $11^\circ$ (Fig.~\ref{fig:azimuth}). The anisotropy of
$\Dtwo$, fitted to the offset-removed GPM-Ku $a_2$, reproduces the positive values
at moderate incidence and their growth with wind to $0.01$--$0.02\dB$ in sample. The amplitudes are tabulated per
$2\,\mathrm{m\,s^{-1}}$ wind class between $4$ and $14\,\mathrm{m\,s^{-1}}$. The
data fix only the products of the closure parameters, namely
$\lambda_3^{0}\sqrt{D_c}$ and the anisotropy ratio $D_{22}/D_{20}=0.022$--$0.037$
at the kernel lag. The month-block interval of this ratio includes zero, so the
data fix the shape of $a_2$ but not a significant anisotropy amplitude.

\begin{figure}[t]
  \centering
  \includegraphics[width=\columnwidth]{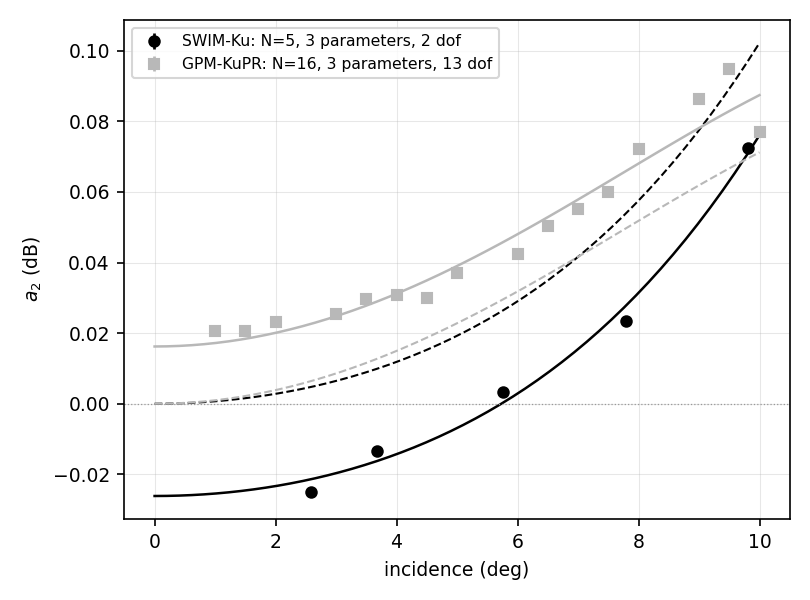}
  \caption{Cross-sensor $a_2$. Markers: SWIM-Ku (dark) and GPM-KuPR (light)
  $a_2(\theta)$ for $6\le\Uten<10\,\mathrm{m\,s^{-1}}$ and $0.5\le\Hs<5$\,m. Solid: fits $c_0+c_2\tan^2\theta+c_4\tan^4\theta$; dashed: the
  geophysical part without $c_0$. The slopes $c_2$ are positive and comparable
  (SWIM $+2.3$, GPM $+3.2$, per unit $\tan^2\theta$), while the offsets $c_0$ have opposite signs
  (SWIM $-0.026$, GPM $+0.016\dB$).}
  \label{fig:directional}
\end{figure}

The near-nadir $a_2$ differs between sensors rather than between bands. GPM-KuPR
and GPM-KaPR $a_2$ are both positive, whereas SWIM $a_2$ is negative near nadir.
The scalar specular model imposes $a_2(0)=0$, so each sensor is fitted with
$a_2(\theta)=c_0+c_2\tan^2\theta+c_4\tan^4\theta$ (Fig.~\ref{fig:directional}). The
nadir-vanishing slope $c_2$ is positive for both, as geometric optics predicts. The
intercept $c_0$, an extrapolation to nadir, is opposite in sign
(Fig.~\ref{fig:directional}). It is therefore treated as a sensor-associated offset,
whose origin in the beam sampling remains to be assessed. The directional coefficients are fitted to the
GPM-Ku harmonics after subtraction of this offset, and a user comparing with SWIM
must first remove its own $c_0=-0.026\dB$.

The Ka harmonics follow from the GPM Ku/Ka ratio, formed after removing the GPM
offset from both bands, on the assumption that it is band-common. Over $8$--$12\,\mathrm{m\,s^{-1}}$ and $2$--$6^\circ$
($12$ common months, $1000$ month-block replicates), the ratios are
$a_1\,\mathrm{Ka/Ku}=1.90$ and $a_2\,\mathrm{Ka/Ku}=2.42$, with month-block
sampling intervals $[1.89,1.92]$ and $[2.38,2.46]$. These intervals exclude the
offset assumption: without the offset removal, $a_2\,\mathrm{Ka/Ku}=2.12$. The
Ka directional model is therefore the Ku model times these two empirical ratios,
which are extrapolated outside the window where they are measured. Assuming equal modulation in both bands would
under-predict Ka roughly twofold.

\section{Relation to Prior Models and Operational Use}
\label{sec:discussion}

\subsection{Comparison with GO4 on equal footing}
\label{sec:disc-go}
\begin{figure*}[t]
  \centering
  \includegraphics[width=\textwidth]{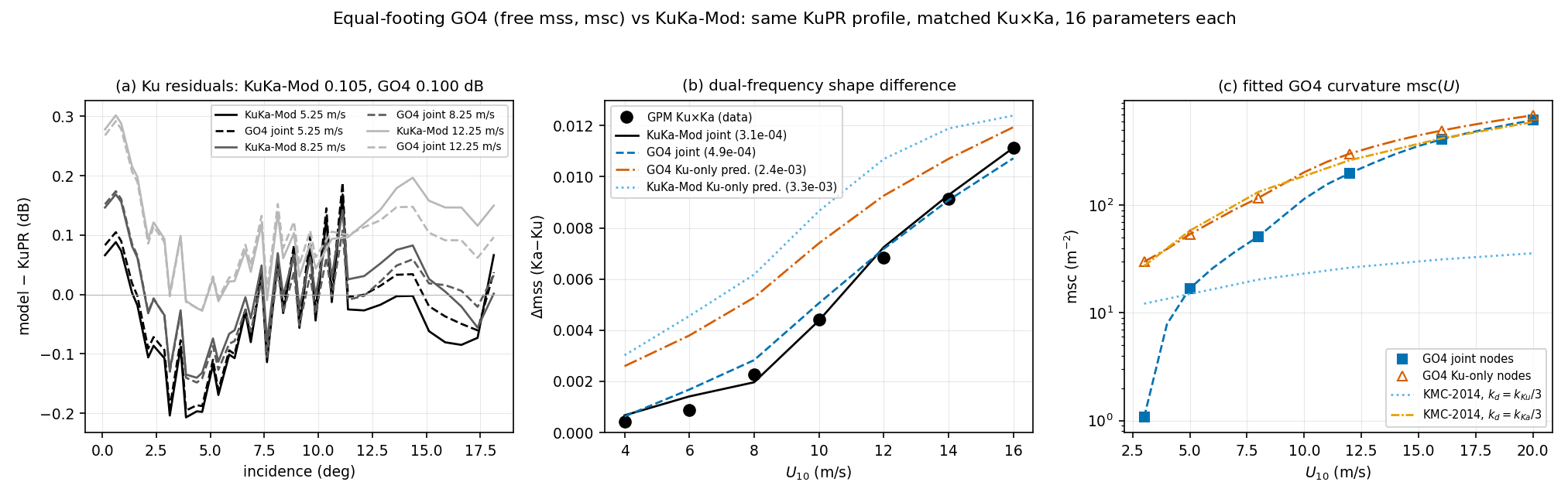}
  \caption{GO4 and KuKa-Mod fitted on equal footing, with the same objective and
  $16$ parameters each. (a)~Ku profile residuals (model minus KuPR) at $5.25$,
  $8.25$ and $12.25\,\mathrm{m\,s^{-1}}$: wRMSD $0.100\dB$ (GO4) and $0.105\dB$
  (KuKa-Mod) over $902$ bins. (b)~The shape difference against wind: matched GPM
  data (markers), KuKa-Mod joint (solid), GO4 joint (dashed), and the Ku-only fits
  of GO4 (dash-dot, $+523\%$ at $4\,\mathrm{m\,s^{-1}}$) and KuKa-Mod (dotted,
  $+626\%$). (c)~The fitted GO4 curvature $\mathrm{msc}(\Uten)$ against the KMC
  spectral msc at the Ku and Ka dividing scales $k_r/3$.}
  \label{fig:go4joint}
\end{figure*}

The closest physical alternative to KuKa-Mod is the GO4 model \citep{boisot2015},
derived in \citet{nouguier2009phd}. GO4 expands the same Kirchhoff integral to fourth order in lag,
where the mean-square curvature msc enters as a finite-wavelength correction
(Part~I). For Gaussian slopes its isotropic form is $\snot_{\rm GO4}=\snot_{\rm GO}[1+\tfrac12A(1-2x+\tfrac12x^2)]$, where
$x=\tan^2\theta/\mss$ and $A=\mathrm{msc}/(\qz^2\mss^2)$. Including slope kurtosis replaces $A$ in the bracket by
$\alpha=\tfrac23\lambda_4+A$, and only the curvature part vanishes as $\qz\to\infty$.
Because $A$ depends on
$\qz$, GO4 predicts a shape difference once $\mathrm{msc}(\Uten)$ is known. For a
fair test, GO4 was fitted with the objective \eqref{eq:objective}, with six $\mss$
and six msc wind nodes and the same level, Bragg flank and weights, that is, with
$16$ parameters (Table~\ref{tab:exp}, row~7; Fig.~\ref{fig:go4joint}).

On the Ku profile the two models tie. GO4 reaches $0.100\dB$ and KuKa-Mod
$0.105\dB$ over the $902$ bins, with $0.110$ and $0.111\dB$ on the
$0$--$11^\circ$ core, so the small GO4 gain lies on the flank. Across bands GO4
does less well. It reproduces the shape difference with an RMSD of
$4.1$--$4.9\times10^{-4}$, against $3.1\times10^{-4}$, and transfers to Ka with a
shape RMSD of $0.25$--$0.26\dB$, against $0.21\dB$, one Ka level each (ranges over the
regularization strengths; row~7 of Table~\ref{tab:exp} is the reference). Fitted to Ku
alone, it fails the withheld-band test as the logarithmic form does ($+523\%$ at
$4\,\mathrm{m\,s^{-1}}$). The need for the second frequency is thus a property of
the data, not of the model family.

The fitted GO4 curvature also behaves as an effective parameter. It rises from an order of magnitude below the KMC msc at the Ku
dividing scale at $3\,\mathrm{m\,s^{-1}}$ to the Ka-scale value at
$20\,\mathrm{m\,s^{-1}}$, while the expansion parameter $A$ at nadir reaches
$0.30$--$0.37$ at $12$--$16\,\mathrm{m\,s^{-1}}$, at the edge of validity of the
truncation. Adding a wind-constant kurtosis brings GO4 to $0.091\dB$ on Ku but
leaves the Ka shape RMSD at $0.23\dB$. The forward equation is thus shared with
GO4. What is new is its inverse use: a centimetre-scale $\Dtwo(r)$ retrieved from
matched Ku$\times$Ka observations, with no imported spectrum and no expansion in lag.

At Ku band, KuKa-Mod overlaps the re-referenced \citet{hossan2021} DPR model and
the refitted KuLMOD (Fig.~\ref{fig:compare}). It differs from them in that the same
parameters predict the Ka profile with one added level, whereas the empirical GMFs
fit each band separately.

\subsection{Operational use}
\label{sec:disc-ops}
These properties make KuKa-Mod a candidate forward model for testing in near-nadir
wind processors. It is referenced to one stable absolute scale, GPM KuPR V07A, to which
published GMFs are adjusted by a constant offset and SWIM by its per-beam
calibration. It is delivered at the GPM footprint scale, and other
footprints are reached by the prescribed sub-footprint correction of
Section~\ref{sec:val-inc}, evaluated per instrument. The parameter set
(Table~\ref{tab:abnodes}), a reference evaluator (Appendix~\ref{app:recipe}) and
Ku/Ka look-up tables on the $(\Uten,\Phi,\theta)$ grid at the reference sea
state are released, with a Ku variant refitted against the stress-equivalent wind
$U_{10S}$ (Section~\ref{sec:val-stab}). The sea-state tilt belongs to the forward model.
Preliminary wind retrievals show that a sea-state table axis lowers the wind error at low
wind, by about $4\%$ at $2$--$4\,\mathrm{m\,s^{-1}}$, but gives no gain above
$8\,\mathrm{m\,s^{-1}}$, because sea state grows with wind and conditioning on it lowers
the wind sensitivity of the table (Section~\ref{sec:val-hs}). A four-dimensional table is
therefore not recommended at high wind, and a separate study is needed to separate the
sea-state and wind effects.

The validity envelope follows from the data. The wind axis is fitted to KuPR over $3$--$14\,\mathrm{m\,s^{-1}}$, constrained by
the matched shape difference to $16\,\mathrm{m\,s^{-1}}$ and extrapolated to
$20\,\mathrm{m\,s^{-1}}$. From $8\,\mathrm{m\,s^{-1}}$ upward the level of the
tables of Section~\ref{sec:val-highwind} should be used. Below $3\,\mathrm{m\,s^{-1}}$ the GMF is unvalidated. C band lies beyond the lags the two kernels constrain and remains an extrapolation, although Ku/C altimeters have recorded it since TOPEX \citep{chapron1995,gourrion2002cband}; the Kirchhoff analysis of that pair gave the initial form of the spectrum of \citet{elfouhaily1997}.

\section{Limitations and Future Work}
\label{sec:disc-limits}
Two near-nadir frequencies do not separate the slope peakedness \citep{chapron2000} from a
scale-dependent roughness,
so the physical core has no independently fitted kurtosis; separating it would need an
attenuation-corrected measurement and a third frequency. C or Envisat RA-2 S band
($3.2$\,GHz) would lengthen the lag lever arm; three-frequency C/Ku/Ka altimetry was
proposed to estimate the sea-state bias \citep{amarouche2001}. The high-wind level of
$b(\Uten)$ is measured rather than predicted, because the source spectrum
over-predicts it. The Ka level keeps a wind-dependent residual, about $20\%$ in
near-nadir wind sensitivity. The $11$--$18^\circ$ flank is an empirical Bragg term.
The sea-state tilt is measured but not calibrated as an operational law: its size in
$\Hs$ depends on the reference wind, its slope-variance efficiency and the
scatterometer collocation cover one SWIM year, KuPR does not separate a law linear in
$\Hs$ from a quadratic one, the retrieval tests rest on one sample each, and at Ka a
band-common tilt places the zero crossing outside the observed one. A term in wave age
or in the slope variance is left to future work. The released core is referenced to the
ERA5 wind. Its stress-equivalent refit rests on two years (2021--2022) of KuPR--scatterometer collocations
and leaves the directional module unchanged. From $8\,\mathrm{m\,s^{-1}}$ upward the
operational level is the empirical table of Section~\ref{sec:val-highwind}, not the
physical model; only its quantile-matched bins are assessed, distributionally, to
$20\,\mathrm{m\,s^{-1}}$. All data lie within $|\mathrm{lat}|\le60^\circ$. A test on data
withheld in time and region, and a wind-retrieval experiment in the CFOSAT/SWIM or
SWOT processing chains, are the next steps.

\section{Conclusion}
\label{sec:conclusion}

KuKa-Mod is the Ku/Ka realization of the structure-function Kirchhoff framework of
Part~I, built for operational wind retrieval. The scale dependence of the elevation
structure function at centimetre lags is retrieved jointly from the GPM KuPR
incidence profile and the same-footprint Ku$\times$Ka shape difference. The fit
reproduces the KuPR $\snot(\theta,\Uten)$ surface to a wRMSD of $0.105\dB$ over $902$ bins and
the shape difference to an RMSD of $3.1\times10^{-4}$, against
$2.6\times10^{-3}$ for the KMC spectrum. The log-slope
grows approximately as the square of the wind speed. A Ku-only fit matches the Ku
profile equally well but over-predicts the shape difference by $626\%$ at
$4\,\mathrm{m\,s^{-1}}$, so one band does not fix the transfer. GO4, fitted on the
same footing, transfers less well to Ka. The physical core reaches the azimuth-averaged Ka profile with no additional shape parameter
and one fitted level, most of which the seawater dielectric and the clear-sky
attenuation explain. The near-nadir Ka wind sensitivity is about $20\%$ too weak. At
Ku band the model places, with no sea-state coefficient, the incidence at which the
response of $\snot$ to $\Hs$ changes sign. In SWIM this response is consistent with the
long-wave slope variance, but its amplitude, measured at about half the geometric-optics
value, remains unresolved. Against
scatterometer winds the Ka--Ku difference holds to within $0.3\dB$ up to
$25\,\mathrm{m\,s^{-1}}$, and a
level correction measured on GPM gives the operational level of both bands. The azimuth harmonics are reproduced once a sensor-dependent near-nadir offset
is removed.

The GMF $\snot(\theta,\Uten,\Phi)$ extends to
$18^\circ$ and is released as a parameter set, an evaluator and look-up tables.

\section*{Data and Code Availability}
The KuKa-Mod v1 record is deposited on Zenodo
(\href{https://doi.org/10.5281/zenodo.24084870}{doi:10.5281/zenodo.24084870}). It contains the parameter file (the $(a,b)$ nodes, the per-band levels
and their temperature reference, the directional amplitudes and Ku/Ka ratios, and
the Bragg triple), a standalone evaluator implementing Appendix~\ref{app:recipe}
with NumPy and SciPy only, and Ku and Ka look-up tables of linear $\snot$ at the
reference sea state on the $(\Uten,\Phi,\theta)$ grid ($0.2$--$50\,\mathrm{m\,s^{-1}}$, $0$--$180^\circ$,
$0$--$18^\circ$), which follow the extrapolation rule of Appendix~\ref{app:recipe}
outside the validity envelope of Section~\ref{sec:disc-ops}, the Ku table of
the core refitted against $U_{10S}$ (Section~\ref{sec:val-stab}), and the high-wind
tables of Section~\ref{sec:val-highwind}: the GPM Ku and Ka reference tables and the
SWIM instrument table, each on the $U_{10S}$ axis and stored with $\Delta_{\rm hw}$ and
the name of its base table.
Validation summaries reproduce Tables~\ref{tab:abnodes} and \ref{tab:compare} and
the joint column of Table~\ref{tab:holdout}; the fitting code and the binned GPM and
SWIM data are not part of the record. Earlier experimental sea-state tables, fitted against the ERA5 wind, are archived
for reproducibility only and are not part of the model or the recommendations of
this paper (Section~\ref{sec:val-hs}). The SWIM recalibration software and
tables are available from the IFREMER Wind and Wave Operation Center (IWWOC) on
demand. GPM DPR V07A products are distributed by NASA/JAXA, CFOSAT/SWIM L2 products
by CNES/AVISO and CNSA, ERA5 by the Copernicus Climate Data Store, and WAVEWATCH~III
is an open-source community model.

\appendices

\section{Evaluation recipe}
\label{app:recipe}
The core model evaluates in five steps, plus an operational level, given wind $\Uten$, incidence $\theta$,
relative azimuth $\Phi$, sea-surface temperature $T$
and band (frequency $f$, hence $k_r=2\pi f/c$).

\begin{enumerate}
\item \emph{Surface parameters.} Interpolate $a(\Uten)$ and $b(\Uten)$ from the
  nodes of Table~\ref{tab:abnodes} with PCHIP in $\ln\Uten$. Outside
  $3$--$20\,\mathrm{m\,s^{-1}}$ continue them linearly in $\ln\Uten$ with the end
  slope, floor them at zero and hold them constant below $1\,\mathrm{m\,s^{-1}}$;
  these ranges are unvalidated. The nodes describe the reference sea state
  $\Hs\in[1.5,3.5]$\,m; no sea-state term is added (Section~\ref{sec:val-hs}).
\item \emph{Structure function.} Form
  $\Dtwo(r)=\tfrac12 r^2[a+b\ln(1+r_{\rm ref}/r)]$ with $r_{\rm ref}=1$\,m
  \eqref{eq:marginallog}.
\item \emph{Level.} Evaluate $R_{\rm eff}^2=R_F^2(f,T,35)\,C_f$ with the
  \citet{meissner2004} permittivity. The fitted levels are referenced to
  $T=18.3^\circ$C: $R_{\rm eff}^2=0.581$ at Ku ($C_f=0.944$) and $0.428$ at Ka,
  so the temperature enters as the ratio $R_F^2(f,T)/R_F^2(f,18.3^\circ\mathrm{C})$.
\item \emph{Specular core.} Evaluate the integral \eqref{eq:hankel} with
  $\qz=2k_r\cos\theta$ and $\qh=2k_r\sin\theta$.
\item \emph{Optional terms.} Multiply by $[1+A_1\cos\Phi+A_2\cos2\Phi]$ from the
  tabulated directional amplitudes; at Ka scale the Ku harmonics by $1.90$ and
  $2.42$. Add the gated Bragg term \eqref{eq:full_bragg}. For a
  coarse footprint, the sub-footprint renormalization \eqref{eq:d2eff} is applied
  outside the released evaluator.
\item \emph{Operational level.} For wind retrieval, evaluate steps~1--5 at
  $U_{10S}$ with the core of Section~\ref{sec:val-stab} and multiply by
  $10^{\Delta_{\rm hw}/10}$ from the high-wind table of the band \eqref{eq:highwind}.
\end{enumerate}
The band is changed by repeating steps~3--5 at the new $k_r$, with no shape refit.
The released evaluator follows this sequence when called without $\Hs$ (its legacy sea-state
option is not part of the model), flooring $a$
at $10^{-5}$ and clipping the output to $[-50,30]\dB$ in extrapolation.

\section{Processing details}
\label{app:proc}
\emph{Screening.} Rain contaminates near-nadir $\snot$, so both
missions are screened for rain, ice, land and coast. DPR rays are kept when the
surface is ocean, the precipitation, saturation and ice flags are clear, the
quality flag is good, $|\mathrm{lat}|\le60^\circ$ and the coast is at least
$30$\,km away. SWIM segments are kept when the sea-ice, shallow-bathymetry and
low-incidence flags are clear, land and ice fractions are zero, the nadir rain flag
is clear, $|\mathrm{lat}|\le60^\circ$ and the coast is at least $30$\,km away. The
SWIM rain flag comes from the nadir track, so the off-nadir beams inherit a
conservative along-track mask.

\emph{SWIM calibration.} SWIM $\snot$ is brought to the KuPR V07A scale by three
fixed corrections: a per-beam monthly drift correction referenced to an
$8\,\mathrm{m\,s^{-1}}$ ERA5-wind timeline, a per-beam azimuthal correction in the
orbital frame, and a per-beam, time-invariant incidence-dependent gain correction
derived from SWIM$\times$GPM collocations. The chain drives the SWIM$-$KuPR
residual flat to a few thousandths of a dB and removes an in-flight per-beam gain
step at the outage resumption of 2022-08-17.

\emph{High-wind level.} The KuKa azimuth harmonics are removed from $\snot$ before
the fit. In each $0.25^\circ$ bin with a rank correlation between $\snot$ and $U_{10S}$
below $-0.5$, the noise of the $\snot$-equivalent wind $x$, obtained through the
azimuth-mean model curve, and of $U_{10S}$ come from triple collocation with ERA5 on the
uniform subsample over $5$--$20\,\mathrm{m\,s^{-1}}$. The weighted $U_{10S}$ quantiles
$u_q$ are rescaled about their mean $\bar u$ as $\bar u+k(u_q-\bar u)$, with
$k^2=(\sigma_u^2-e_u^2+e_x^2)/\sigma_u^2$, and $\Delta_{\rm hw}$ at that wind is the
$\snot$ quantile of order $1-q$ minus the model. Quantiles with fewer than $10^3$
samples, in weight, beyond them are not matched, since a rare dark population, such as
unflagged rain, would otherwise be mapped onto the strongest winds. Bins that are
wind-blind, weakly correlated, too noisy ($e_x>2.5\,\mathrm{m\,s^{-1}}$) or with a
degenerate collocation use weighted medians in $0.5\,\mathrm{m\,s^{-1}}$ wind bins
holding at least $200$ samples in weight. $\Delta_{\rm hw}$ is then smoothed by a
penalized spline in wind, held constant outside the support, blended from zero over
$4$--$8\,\mathrm{m\,s^{-1}}$, smoothed across incidence with weights $\sqrt n$, and
constrained so that the azimuth-mean $\snot$ does not increase with wind above
$5\,\mathrm{m\,s^{-1}}$ up to $9.5^\circ$.

\emph{Dielectric, frequencies and numerics.} The Ku Fresnel reflectivity varies by
$0.18\dB$ between $-2$ and $30^\circ$C, hence the temperature reference of the
levels. The model uses $13.575$ and $35.75$\,GHz; at the DPR frequencies $\snot$
changes by at most $0.008\dB$. The integral \eqref{eq:hankel} uses $200$
Gauss--Legendre nodes and is converged to well below $10^{-3}\dB$.

\emph{Realizability.} The logarithmic $\Dtwo$ is a local closure over the radar
lags and does not saturate. A Gaussian-subordination continuation has a spectrum
negative only in a shallow dip of about $1\%$ of the peak, and a globally
positive-definite continuation has not been constructed (Part~I). The regularized $D_3$ closure gives a
standardized skewness of at most $0.16$ at the Ku kernel lag and at most $1.24$ as
$r\to0$, inside the Pearson bound $\sqrt2$ at Gaussian kurtosis.

\section*{Acknowledgment}
This work was performed as a part of the CFOSAT IWWOC development. The IWWOC is co-funded by CNES and IFREMER. The authors
thank the CFOSAT/SWIM and GPM/DPR mission teams and data providers.

\bibliographystyle{IEEEtranN}
\bibliography{refs}

\end{document}